\documentclass[%
 preprint,
 amsmath,amssymb,
 aps,
prb,
]{revtex4-2}

\usepackage{graphicx}
\usepackage{dcolumn}
\usepackage{bm}
\usepackage{xcolor}

\begin{document}
\newcommand\TrnRotOpe[2]{ \mathcal{R}_{\left\{ #1 | #2 \right\}} }
\newcommand\TrnRot[2]{ \left\{ #1 | #2 \right\} }
\newcommand\DeltaT[3]{ \boldsymbol{\Delta}_{\TrnRot{#1}{#2},#3}}
\newcommand\tr[1]{ \mathrm{tr}\left( #1 \right) }
\newcommand\Tr[1]{ \mathrm{Tr}\left( #1 \right) }
\newcommand\LowdinOrb[1]{ u_{#1}}
\def\unitmatrix{1}
\def\density{\rho}
\def\IRBZ{\mathrm{IRBZ}}
\def\FBZ{\mathrm{FBZ}}
\def\Hxc{\mathrm{Hxc}}
\def\ppnl{\mathrm{ppnl}}
\def\DFTplusUgroup{\mathcal{I}}
\def\KS{\mathrm{KS}}
\def\CN{\mathrm{CN}}
\def\ext{\mathrm{ext}}
\def\eff{\mathrm{eff}}
\def\ind{\mathrm{ind}}
\def\Nbar{\overline{N}}
\def\Ubar{\overline{U}}
\def\Jbar{\overline{J}}
\def\calM{\mathcal{M}}
\def\calP{\mathcal{P}}
\def\calU{\mathcal{U}}
\newcommand\Projection[1]{\mathcal{P}^{#1}}
\newcommand\OrbitalGroup[1]{\mathscr{O}^{#1}}
\newcommand\locorb[2]{\phi^{#1}_{#2}}
\def\Overlap{O}
\def\DFTplusU{\mathrm{DFT+U}}
\def\VDFTplusU{V^{\DFTplusU}}
\newcommand\Commutator[2]{ \left[ #1 , #2 \right] }
\newcommand\Ifrac[2]{ \mathop{} \left. #1 \right/ #2 }
\def\Nabla{\boldsymbol{\nabla}}
\def\bfzero{\boldsymbol{0}}
\def\bfone{\boldsymbol{1}}
\def\calI{\mathcal{I}}
\def\bfa{\mathbf{a}}
\def\bfc{\mathbf{c}}
\def\bfj{\mathbf{j}}
\def\bftau{\boldsymbol{\tau}}
\def\bfb{\mathbf{b}}
\def\bfp{\mathbf{p}}
\def\bfn{\mathbf{n}}
\def\bfk{\mathbf{k}}
\def\bfr{\mathbf{r}}
\def\bfv{\mathbf{v}}
\def\bfx{\mathbf{x}}
\def\bfy{\mathbf{y}}
\def\bfR{\mathbf{R}}
\def\bfA{\mathbf{A}}
\def\bfE{\mathbf{E}}
\def\bfT{\mathbf{T}}
\def\CrysSymGroup{\mathfrak{G}}
\def\CrysSymSubGroup{\mathcal{G}}

\preprint{APS/123-QED}

\title{Development of \textit{ab initio} Hubbard parameter calculation schemes in the $\mathbf{k}$-point sampling real-time TDDFT program in CP2K}

\author{Kota Hanasaki}
\email{kota.hanasaki@chem.uzh.ch}
\author{Sandra Luber}%
\affiliation{Department of Chemistry, University of Zurich, 8057 Zurich, Switzerland}

\date{\today}

\begin{abstract}
We implemented ab initio Hubbard parameter calculation schemes in the $\bfk$-point
sampling real-time TDDFT (RT-TDDFT) program in CP2K. We propose a new linear-response-based calculation scheme for energy-dependent Hubbard parameters. Our scheme extends the minimum-tracking linear-response method proposed in [Moynihan et al., arXiv preprint arXiv:1704.08076(2017); E. B. Linscott et al., Phys. Rev. B 98, 235157 (2018)] to realize the calculation of energy-dependent Hubbard parameters that reflect the exchange-correlation (xc) effects included in the xc-functional.

We discuss the properties of the minimum-tracking linear-response method in comparison to another promising scheme, ACBN0 [Agapito et al., Phys. Rev. X, 5, 011006 (2015)]. We show that, while neither clearly outperforms the other in the accuracy of static property calculations, each has a distinct dynamical application depending on its theoretical formulation.
\end{abstract}
\maketitle

\section{Introduction}\label{sec:Introduction}
The density functional theory (DFT)~\cite{Hohenberg-Kohn,Kohn-Sham} and its time-dependent (TD) extensions~\cite{Runge-Gross}, are undoubtedly vital tools for \textit{ab initio} calculations of material properties and dynamics. 
However, it has long been known that DFT with local/semilocal exchange-correlation (xc) functionals fails in the calculations of materials with strong electron-electron correlations among localized $d$ or $f$ orbitals, such as transition metal (TM) oxides. 
This problem is known to arise from the self-interaction errors~\cite{PZ,Mori-Cohen-WeitaoYang_self-interacton_error_SIE_JCP2006,Mori-Sanchez2008Self-interaction_delocalization,Cococcioni_DFT+Ulr_PRB2005,Moynihan-ORegan2017self-consistentU,Burgess-Linscott-ORegan_BLOR_flat-plane} in local/semilocal xc functionals. 
DFT+U~\cite{Anisimov-Zaanen-Andersen_DFT+U,Dudarev}, which introduces an explicit on-site interaction among correlated orbitals, provides a cost-efficient solution to the self-interaction problem. 
Such cost-efficiency is vital for high-throughput computations~\cite{extensive_use_of_DFT+U_MDPI2021,High-throughput_HubbardU_PhysRevMat} for material design of correlated materials~\cite{Correlated_materials_designAdler_2019_Rep._Prog._Phys._82_012504}, catalysts containing transition metal oxides~\cite{DFT+U_catalytic-materials}, etc.

One of the most critical factors in DFT+U calculations is the values of the effective on-site interaction parameters, which are commonly referred to as `Hubbard $U$' and 'Hund $J$'. For simplicity of notation, we hereafter call them 'Hubbard parameters' in a collective sense.
An empirical approach is to fit the values of the Hubbard parameters so that the calculation reproduces some experimentally observed physical quantities, such as the band gaps. 
However, it has been known that the most appropriate values of the Hubbard parameters in general depend on the choice of physical quantity to be fitted~\cite{DFT+U_catalytic-materials}. It is also known that those values sensitively depend on the chemical environment, such as ligands and their configuration, oxidation number of the TM atoms, etc.~\cite{DFT+U_catalytic-materials,Local_environment_dependent_GGA+U}. Researchers have hence developed more consistent \textit{ab initio} approaches for the derivation of Hubbard parameters. 

Theoretically, the Hubbard parameters are characterized as spherically averaged screened interactions among correlated orbitals. It therefore depends on the orbital energy, or, more precisely, in terms of many-body theory, the energy of quasiparticles. 
The excited-state properties in the energy range far from the Fermi level should therefore be calculated using energy-dependent Hubbard parameters. 
Indeed, such energy dependence of the Hubbard parameters was found to affect the computed properties~\cite{Aryasetiawan-Imada2004_HubbardU_cRPA,Marzari_DFT+Uw_BiFeO3_arXiv2025}.
In non-equilibrium dynamics, such as laser-induced dynamics, the Hubbard parameters depend on time, reflecting the dynamics of screening electrons~\cite{Rubio_tdDFT_ACBN0_2017,Rubio_MoTe2_LifshitzTransition,Rubio_attosecond_response_RT-TDDFT,Eckstein_DMFT_HubbardU_renormalization}.
Those facts further motivate us to calculate the Hubbard parameters using \textit{ab initio} methods, possibly in an energy- or time- dependent manner. This is the main focus of this paper.

There are a variety of Hubbard parameter calculation techniques, including the constrained local density approximation (cLDA) method~\cite{Anisimov-Gunnarson_constrainedLDA_calc_U_DFt+U_PhysRevB.43.7570}, the constrained random-phase approximation (cRPA) method~\cite{Aryasetiawan-Imada2004_HubbardU_cRPA}, linear response methods~\cite{Cococcioni_DFT+Ulr_PRB2005,Moynihan-ORegan2017self-consistentU,Linscott-ORegan_MTLR_PRB2018}, and a mean-field-like (in this paper, we refer to ACBN0 as mean-field-like in the sense that the Hubbard parameters are computed using only the static correlation effects incorporated in the density matrix and the Kohn-Sham orbitals) approach in the ACBN0 quasi-hybrid functional~\cite{Agapito-Nardelli2015ACBN0}.
Among them, cRPA has an apparent advantage of being able to calculate energy-dependent Hubbard parameters, whereas it has been pointed out~\cite{Werner_limitations_cRPA_PhysRevB.98.235151,Carta-Timrov_Bridging_cRPA_and_LR_2505.03698v1} that in some cases, it overestimates the screening due to missing terms in the random-phase approximation (RPA)~\cite{Werner_limitations_cRPA_PhysRevB.98.235151} or hybridization of \textit{screening} and \textit{correlated} orbitals~\cite{Carta-Timrov_Bridging_cRPA_and_LR_2505.03698v1} (here, the \textit{correlated} orbitals mean the orbitals to which we introduce Hubbard-like on-site interactions in the DFT+U scheme, whereas the \textit{screening} orbitals refer to all other orbitals in the calculation). 
On the other hand, the linear response schemes, whose applications have so far focused on the calculation of static (energy-independent) Hubbard parameters, are built on the linear-response theory and the analysis of self-interaction associated with a given xc functional. The latter point is in contrast to cRPA, where the screened interaction is calculated with only the direct Coulombic interaction (in RPA) taken into account. Carta \textit{et al.} in Ref.~\cite{Carta-Timrov_Bridging_cRPA_and_LR_2505.03698v1}, reported higher robustness of the linear response scheme in the cases where cRPA fails. 
It is therefore expected that a possible linear-response-based approach for dynamical Hubbard parameters should work as an alternative to cRPA in such difficult cases.

A remarkable advantage of the mean-field-like treatment in the ACBN0 quasi-hybrid functional~\cite{Agapito-Nardelli2015ACBN0} is in its applicability to non-equilibrium dynamics~\cite{Tancogne-Dejean_Rubio_NiO_2020,Rubio_MoTe2_LifshitzTransition,Rubio_attosecond_response_RT-TDDFT}. Tancogne-Dejean~\textit{et al.}~\cite{Tancogne-Dejean_Rubio_NiO_2020}, Beaulieu~\textit{et al.}~\cite{Rubio_MoTe2_LifshitzTransition}, and Cazali~\textit{et al.}~\cite{Rubio_attosecond_response_RT-TDDFT} applied 
RT-TDDFT+U with time-dependent Hubbard parameters calculated using ACBN0 
to femto to attosecond field-induced dynamics and analyzed its resultant band-gap renormalization effects. 
However, as we discuss later, despite its successful applications to static~\cite{Agapito-Nardelli2015ACBN0,Rubio_tdDFT_ACBN0_2017} and dynamical~\cite{Tancogne-Dejean_Rubio_NiO_2020,Rubio_MoTe2_LifshitzTransition,Rubio_attosecond_response_RT-TDDFT} problems, the fact that the expressions of Hubbard parameters in ACBN0 are not derived from standard many-body theory makes it difficult to analyze or predict their behaviors. 

In this paper, we report the implementation of the minimum-tracking linear-response method~\cite{Moynihan-ORegan2017self-consistentU,Linscott-ORegan_MTLR_PRB2018} and the ACBN0 quasi-hybrid functional~\cite{Agapito-Nardelli2015ACBN0} into our recently developed \textbf{k}-point-sampling real-time TDDFT (RT-TDDFT) framework~\cite{HanasakiLuber,CP2Ksoftwarepaper} built on the CP2K software suite~\cite{CP2K}. We discuss the advantages and limitations of these \textit{ab initio} Hubbard parameter calculation methods. We also work on an extension of the linear-response theory to compute energy-dependent Hubbard parameters using RT-TDDFT.

This paper is organized as follows. In Sec.~\ref{sec:Formulation}, 
we briefly review the theory and show our implementation of the Hubbard parameters computation schemes in RT-TDDFT.
In Sec.~\ref{sec:CalculationResults}, we show our calculation results. Section~\ref{sec:Summary} is devoted to the summary and discussions. 


\section{Formulation} \label{sec:Formulation}
We briefly review the theory and show our implementation of RT-TDDFT (Sec.~\ref{subsec:DFTandRT-TDDFT}), DFT+U (Sec.~\ref{subsec:DFT+U}), Hubbard parameter calculation schemes (Sec.~\ref{subsec:AbInitioCalcOfHubbardParameters}), and energy-dependent Hubbard parameters (Sec.~\ref{subsec:Energy-dependent-U}). 
The details of the first two were already shown in our previous publication~\cite{HanasakiLuber}, whereas we show the essential parts for the convenience of later discussions. The latter two are the main topic in this paper, including our new formulation in Sec.~\ref{subsubsec:Extension_of_LR_schemes}. 
\subsection{DFT and RT-TDDFT}\label{subsec:DFTandRT-TDDFT}
The Kohn-Sham (KS) equation for a periodic system is solved as
\begin{align}
    H^{KS}_\sigma \psi_{\lambda\bfk\sigma}(\bfr)=\varepsilon_{\lambda\bfk\sigma}\psi_{\lambda\bfk\sigma}(\bfr)
\end{align}
where $\sigma$ indicates the spin projection, $H^{KS}_\sigma$ represents the KS Hamiltonian, and $\psi_{\lambda\bfk\sigma}$ is a KS orbital characterized by the Bloch vector $\bfk$ and orbital index $\lambda$.
KS orbitals are expanded in $\bfk$-adapted basis functions $\{\chi_{\mu\bfk}(\bfr)\}$ as
\begin{align}
    \psi_{\lambda\bfk\sigma}(\bfr) = \sum_\mu \chi_{\mu\bfk}(\bfr) C^{\mu}_{\lambda\bfk\sigma}
\end{align}
with $C^{\mu}_{\lambda\bfk\sigma}$ being coefficients.
In CP2K, the $\bfk$-adapted basis functions are formally constructed from Gaussian atomic orbitals (AOs) $\{\chi_\mu(\bfr)\}$ as
\begin{align}
    \chi_{\mu\bfk}(\bfr) = \sqrt{\frac{1}{N_L}} \sum_{\bfT} \chi_\mu(\bfr-\bfT) e^{i\bfk\cdot\bfT}
    \label{eqn:k-adaptedAO}
\end{align}
where $\bfT$ represents the lattice vector, $N_L$ represents the number of unit cells or, equivalently, the number of $\bfk$-points in the first Brillouin zone.
Properties of the system are calculated from the density matrix
\begin{align}
    \rho_\sigma(\bfr,\bfr')&=\sum_{\lambda\bfk}\psi_{\lambda\bfk\sigma}(\bfr) \theta_{\lambda\bfk}\psi^\ast_{\lambda\bfk\sigma}(\bfr') \nonumber\\
    &= \sum_\bfk \sum_{\mu\nu} \chi_{\mu\bfk}(\bfr) D^{\mu\nu}_{\bfk\sigma} \chi^{\ast}_{\nu\bfk}(\bfr'),
\end{align}
where $\theta_{\bfk\lambda}$ represents the occupation number of the KS orbital $\psi_{\lambda\bfk}$ and 
$D^{\mu\nu}_{\bfk\sigma}$ is the density matrix in the AO representation
\begin{align}
    D^{\mu\nu}_{\bfk\sigma} \equiv \sum_{\lambda} C^{\mu}_{\lambda\bfk}\theta_{\lambda\bfk}C^{\nu\;\ast}_{\lambda\bfk}.
\end{align}

In RT-TDDFT, the time-dependent KS orbitals are propagated by solving the time-dependent KS equation
\begin{align}
    i\hbar \frac{\partial}{\partial t}\psi_{\lambda\bfk\sigma}(\bfr,t)
    = H^{KS}_\sigma(t) \psi_{\lambda\bfk\sigma}(\bfr,t)
\end{align}
with the boundary condition $\psi_{\lambda\bfk\sigma}(\bfr,0)=\psi^{GS}_{\lambda\bfk\sigma}(\bfr)$ where $\psi^{GS}_{\lambda\bfk\sigma}(\bfr)$ is the corresponding eigenfunction of the ground-state KS Hamiltonian.
In the linear response RT-TDDFT calculations, which we discuss in Secs.~\ref{subsubsec:Extension_of_LR_schemes} and~\ref{subsec:RSL_MTLR_dynamical}, the ground state KS orbitals at $t=-0$ are perturbed with a perturbation operator including $\delta(t)$ to make the initial state at $t=+0$.


\subsection{DFT+U}\label{subsec:DFT+U}
\subsubsection{The energy functional}\label{subsubsec:EnergyFunctional}
In DFT+U, the Coulomb interactions among selected sets of localized orbitals with strong correlation, hereafter referred to as \textit{correlated orbitals}, are treated explicitly in the form of on-site interactions. 
The set of correlated orbitals, hereafter labeled by symbol $\calI$, is characterized by the atomic species and the orbital angular momentum $\ell_{\calI}$ (e.g., $d$-orbitals of Ni).  
The energy functional is written as
\begin{align}
    E_{\mathrm{DFT+U}}[ n_\uparrow(\bfr), n_\downarrow(\bfr);\{n_{m\sigma}^{(\calI,A)}\}] = E_{\mathrm{DFT}}[ n_\uparrow(\bfr), n_\downarrow(\bfr)]+ E_{\mathrm{Hub}}[\{n_{m\sigma}^{(\calI,A)}\}]
    -E_{\mathrm{DC}}[\{n_{m\sigma}^{(\calI,A)}\}]
\end{align}
where $n_\sigma(\bfr)$ is the density, $n_{m\sigma}^{(\calI,A)}$ is the occupation number of the $m$th orbital in the $A$th atom in the orbital set $\calI$, while $E_{\mathrm{DFT}}$, $E_{\mathrm{Hub}}$, and $E_{\mathrm{DC}}$ are the original DFT energy functional, energy of the on-site interactions, and the double-counting correction term, respectively.
We follow the formulation by Dudarev \textit{et al}~\cite{Dudarev}. The expression of $E_{\mathrm{Hub}}$ is given as
\begin{align}
    E_{\mathrm{Hub}}[\{n_{m\sigma}^{(\calI,A)}\}] = \sum_{\calI} \left( 
    \frac{ \overline{U}^{\calI}-\overline{J}^{\calI}}{2}\sum_{m\neq m',\sigma} n_{m\sigma}^{(\calI,A)} n_{m'\sigma}^{(\calI,A)}
    + \frac{ \overline{U}^{\calI}}{2} \sum_{m,m',\sigma} n_{m\sigma}^{(\calI,A)} n_{m'-\sigma}^{(\calI,A)} \right)
\end{align}
with $\overline{U}^{\calI}$ and $\overline{J}^{\calI}$ representing the spherically averaged Coulomb repulsion and Hund's-rule coupling parameters, respectively,  
and for $E_{\mathrm{DC}}$,
\begin{align}
    E_{\mathrm{DC}}[\{n_{m\sigma}^{(\calI,A)}\}] = \sum_{\calI} \left(
    \frac{ \overline{U}^{\calI}-\overline{J}^{\calI}}{2}N^{(\calI,A)}_\sigma \left( N^{(\calI,A)}_\sigma-1\right) 
    + \frac{ \overline{U}^{\calI}}{2} N^{(\calI,A)}_\sigma N^{(\calI,A)}_{-\sigma}
    \right)
\end{align}
with $N^{(\calI,A)}_\sigma\equiv \sum_m n_{m\sigma}^{(\calI,A)}$. 
The resultant equation is expressed in a more generalized form using the projected density matrix $\rho^{(\calI,A)}_\sigma$ as~\cite{Dudarev}
\begin{align}
E_{\mathrm{DFT+U}}=E_{\mathrm{DFT}} + \sum_{(\DFTplusUgroup,A)}\frac{\overline{U}^{\DFTplusUgroup}-\overline{J}^{\DFTplusUgroup}}{2}\sum_\sigma \left( \tr{ \rho_\sigma^{(\DFTplusUgroup,A)}}-\tr{ \rho_\sigma^{(\DFTplusUgroup,A)} \rho_\sigma^{(\DFTplusUgroup,A)}} \right). 
\label{eqn:DFT+U.rho}
\end{align}
In this scheme, the correction term for each set $\calI$ is characterized by a single effective interaction parameter $\Ubar^{\calI}_{\eff}\equiv \overline{U}^{\DFTplusUgroup}-\overline{J}^{\DFTplusUgroup}$.

\subsubsection{Projection scheme}\label{subsubsec:ProjectionScheme}
The projected density matrix in Eq.~\eqref{eqn:DFT+U.rho} is given as
\begin{align}
    \rho_\sigma^{(\calI,A)}=\calP^{(\calI,A)}\rho_\sigma \calP^{(\calI,A)} 
    \label{eqn:DMprojection}
\end{align}
where the projection operator $\calP^{(\calI,A)}$ in the coordinate representation is given as~\cite{DFTplusU_ORegan2011_SubspaceRep}
\begin{align}
    \calP^{(\calI,A)}(\bfr,\bfr')=\sum_{m,m'\in(\calI,A)} \phi^{(\calI,A)}_m(\bfr) \left(O_{(\calI,A)}^{-1}\right)_{mm'} \phi^{(\calI,A)\;\ast}_{m'}(\bfr')
    \label{eqn:ProjectionOperator}
\end{align}
with $\{\phi^{(\calI,A)}_m(\bfr)\}$ being a set of local orbitals representing the subspace $(\calI,A)$, and $O_{(\calI,A)}$ being their overlap matrix.
We follow the formulation of Chai~\textit{et al.}~\cite{Ziwei} and use the atomic orbitals of an isolated atom for the orbital set $\{\phi^{(\calI,A)}_m(\bfr)\}$. 
They are precomputed as the solution of the ground-state atomic DFT using the same basis set as~\cite{Ziwei}
\begin{align}
    \phi^{(\calI,A)}_m(\bfr)=\sum_{n'} \chi_{(An',\ell_\calI,m)}(\bfr-\bfT_A) c^{(\calI)}_{n'}
    \label{eqn:atomicKSO}
\end{align}
where $\chi_{(An',\ell_\calI,m)}(\bfr)$ is the $n'$th atomic orbital in the atomic species $A$ with angular momentum $\ell_{\calI}$ and the azimuthal quantum number $m$, $\bfT_A$ represents the unit cell coordinate of the $A$th atom, and $c^{(\calI)}_{n'}$ is its associated coefficient. In this scheme, the localized orbitals $\{\phi^{(\calI,A)}_m(\bfr)\}$ form an orthonormal set of real-valued orbitals, hence we hereafter omit the inverse overlap matrix $\left(O_{(\calI,A)}^{-1}\right)_{mm'}$ appearing in Eq.~\eqref{eqn:ProjectionOperator} and use $\calP^{(\calI,A)}(\bfr,\bfr')=\sum_{m\in(\calI,A)} \phi^{(\calI,A)}_m(\bfr) \phi^{(\calI,A)}_{m}(\bfr')$.
Validity of this projection scheme in $\bfk$-point sampling DFT and RT-TDDFT has been shown in Ref.~\cite{HanasakiLuber}.
For the convenience of later discussion, we show the overlap of local orbital $\phi^{(\calI,A)}_m(\bfr)$ and $\bfk$-adapted AO;
\begin{align}
    \langle \chi_{\mu\bfk}|\phi^{(\calI,A)}_m \rangle 
    &= \sqrt{\frac{1}{N_L}}\sum_\bfT e^{-i\bfk\cdot\bfT} \int d^3\bfr \chi^{\ast}_{\mu}(\bfr-\bfT) \sum_{n'} \chi_{(An',\ell_\calI,m)}(\bfr-\bfT_A) c^{(\calI)}_{n'} \nonumber\\
    &= \sqrt{\frac{1}{N_L}} e^{-i\bfk\cdot\bfT_A} \sum_{n'}S^{\mu(An'\ell_\calI m)}_\bfk c^{(\calI)}_{n'},
    \label{eqn:chiphioverlap}
\end{align}
where matrix $S_\bfk$ represents the AO overlap matrix, $S_\bfk^{\mu\nu}\equiv \langle \chi_{\mu\bfk}|\chi_{\nu\bfk} \rangle$.

\subsection{Ab initio calculation of the Hubbard parameters}\label{subsec:AbInitioCalcOfHubbardParameters}
As we discussed in the introduction, we examine two distinct schemes for \textit{ab initio} calculation of the 
Hubbard parameters 
and show our implementation of each scheme in CP2K.

\subsubsection{ACBN0}\label{subsubsec:ACBN0}
Following Agapito~\textit{et al.}~\cite{Agapito-Nardelli2015ACBN0}, we introduce the `renormalized' occupation number. To make it consistent with our projection scheme Eq.~\eqref{eqn:DMprojection}, we use the following expression
\begin{align}
    \Nbar^{(\calI)\;\sigma}_{\psi_{\lambda\bfk\sigma}}\equiv \sum_A \Tr{ \calP^{(\calI,A)}\rho_{\lambda\bfk\sigma}\calP^{(\calI,A)} }
    \label{eqn:Nbar}
\end{align}
where $\rho_{\lambda\bfk\sigma}$ is a part of the density matrix that consists only of a single KS orbital $\psi_{\lambda\bfk\sigma}$ as
\begin{align}
    \rho_{\lambda\bfk\sigma}(\bfr,\bfr')\equiv \psi_{\lambda\bfk}(\bfr)\theta_{\lambda\bfk}  \psi^{\ast}_{\lambda\bfk}(\bfr')
    = \sum_{\mu\nu} \chi_{\mu\bfk}(\bfr) C^{\mu}_{\lambda\bfk\sigma} \theta_{\lambda\bfk\sigma} C^{\nu\;\ast}_{\lambda\bfk\sigma} \chi^{\ast}_{\nu\bfk}(\bfr')
    \label{eqn:rho_lambdabfk}
\end{align}
Equation~\eqref{eqn:Nbar} is therefore evaluated as
\begin{align}
     \Nbar^{(\calI)\;\sigma}_{\psi_{\lambda\bfk\sigma}} &=\sum_A \sum_m \langle \phi^{(\calI,A)}_m|\rho_{\lambda\bfk\sigma}|\phi^{(\calI,A)}_m \rangle \nonumber\\
     &= \sum_A \sum_m \langle \phi^{(\calI,A)}_m|\psi_{\lambda\bfk\sigma}\rangle 
      \theta_{\lambda\bfk\sigma} \langle \psi_{\lambda\bfk\sigma}|\phi^{(\calI,A)}_m\rangle.
      \label{eqn:Nbar2}
\end{align}
We can use Eq.~\eqref{eqn:chiphioverlap} to obtain the overlap $\langle \phi^{(\calI,A)}_m|\psi_{\lambda\bfk\sigma}\rangle$ as
\begin{align}
   \langle \phi^{(\calI,A)}_m|\psi_{\lambda\bfk\sigma}\rangle = \sqrt{\frac{1}{N_L}} \sum_{n,\mu} c^{(\calI)}_n S_\bfk^{(An\ell_\calI m)\mu} C^\mu_{\lambda\bfk\sigma} e^{i\bfk\cdot\bfT_A},
   \label{eqn:phipsioverlap}
\end{align}
hence Eq.~\eqref{eqn:Nbar} is calculated as
\begin{align}
    \Nbar^{(\calI)\;\sigma}_{\psi_{\lambda\bfk\sigma}} &= \sum_A \sum_m \frac{1}{N_L} \left| \sum_{n,\mu} c^{(\calI)}_n S^{(An\ell_\calI m)\mu}_\bfk C^\mu_{\lambda\bfk\sigma} \right|^2 \theta_{\lambda\bfk\sigma} \nonumber\\
    &= \sum_{\mbox{\scriptsize $A$ in unit cell}} \sum_m \left| \sum_{n,\mu} c^{(\calI)}_n S^{(An\ell_\calI m)\mu}_\bfk C^\mu_{\lambda\bfk\sigma} \right|^2 \theta_{\lambda\bfk\sigma}
\end{align}

We then follow Ref.~\cite{Agapito-Nardelli2015ACBN0} to compute the renormalized density matrix $\overline{D}_\sigma$ in the local orbital representation. 
We first work on the coordinate representation to get the expression of the renormalized density matrix as
\begin{align}
    \overline{\rho}^{\calI}_\sigma (\bfr,\bfr')\equiv \sum_{\lambda\bfk} \psi_{\lambda\bfk\sigma}(\bfr) \Nbar^{(\calI)\;\sigma}_{\psi_{\lambda\bfk\sigma}} \psi^{\ast}_{\lambda\bfk\sigma}(\bfr') 
    \label{eqn:renormalizedDM_rRep}
\end{align}
and calculate its local orbital representation as
\begin{align}
    \left( \overline{D}^{(\calI,A)}_\sigma \right)^{mm'} &= \langle \phi^{(\calI,A)}_m | \overline{\rho}^{\calI}_\sigma | \phi^{(\calI,A)}_{m'} \rangle \nonumber\\
    &= \sum_{\lambda\bfk} \langle  \phi^{(\calI,A)}_m | \psi_{\lambda\bfk\sigma} \rangle \Nbar^{(\calI)\;\sigma}_{\psi_{\lambda\bfk\sigma}} \langle \psi_{\lambda\bfk\sigma} |  \phi^{(\calI,A)}_{m'} \rangle.
    \label{eqn:renormalizedDM_AORep}
\end{align}
Using Eq.~\eqref{eqn:phipsioverlap}, we can evaluate Eq.~\eqref{eqn:renormalizedDM_AORep} as
\begin{align}
     \left( \overline{D}^{(\calI,A)}_\sigma \right)^{mm'} &= \frac{1}{N_L} \sum_{\lambda\bfk} \left( \sum_{n,\mu} c^{(\calI)}_n S^{(An\ell_\calI m)\mu}_\bfk C^\mu_{\lambda\bfk\sigma} \right) \Nbar^{(\calI)\;\sigma}_{\psi_{\lambda\bfk\sigma}}
     \left( \sum_{n,\mu} c^{(\calI)}_n S^{(An\ell_\calI m')\mu}_\bfk C^\mu_{\lambda\bfk\sigma} \right)^{\ast}
\end{align}
The averaged interaction parameters $\Ubar$ and $\Jbar$ are evaluated as
\begin{subequations}
{\small
\begin{align}
    \Ubar^{(\calI,A)}=\frac{ \sum\limits_{\substack{m,m',\\m'',m'''}} \left(
   \sum\limits_\sigma \left( \overline{D}^{(\DFTplusUgroup,A)}_{\sigma}\right)^{m m'}
               \left( \overline{D}^{(\DFTplusUgroup,A)}_{\sigma}\right)^{m'' m'''}
   \!\!\!\!\!\!\!+ \sum\limits_\sigma \left( \overline{D}^{(\DFTplusUgroup,A)}_{\sigma}\right)^{m m'}
                 \left( \overline{D}^{(\DFTplusUgroup,A)}_{-\sigma}\right)^{m'' m'''}
   \right) \langle \phi^{(\DFTplusUgroup,A)}_{m'} \phi^{(\DFTplusUgroup,A)}_{m} || \phi^{(\DFTplusUgroup,A)}_{m'''}  \phi^{(\DFTplusUgroup,A)}_{m''}  \rangle 
}{ \sum\limits_{m\neq m'\;\sigma} \left(N^{(\DFTplusUgroup,A)}\right)_m^\sigma 
\left(N^{(\DFTplusUgroup,A)}\right)_{m'}^\sigma
   + \sum\limits_{m,m'\;\sigma}
\left(N^{(\DFTplusUgroup,A)}\right)_m^\sigma
\left(N^{(\DFTplusUgroup,A)}\right)_{m'}^{-\sigma}}
\label{eqn:Ubar_ACBN0}
\end{align}
}
and
\begin{align}
    \Jbar^{(\calI,A)}=\frac{ \sum\limits_{\substack{m,m',\\m'',m'''}} \left(
   \sum\limits_\sigma \left( \overline{D}^{(\DFTplusUgroup,A)}_{\sigma}\right)^{m m'}
               \left( \overline{D}^{(\DFTplusUgroup,A)}_{\sigma}\right)^{m'' m'''}
   \right) \langle \phi^{(\DFTplusUgroup,A)}_{m'} \phi^{(\DFTplusUgroup,A)}_{m''} || \phi^{(\DFTplusUgroup,A)}_{m'''}  \phi^{(\DFTplusUgroup,A)}_{m}  \rangle
}{ \sum\limits_{m\neq m'\;\sigma}
\left(N^{(\DFTplusUgroup,A)}\right)_m^\sigma
\left(N^{(\DFTplusUgroup,A)}\right)_{m'}^\sigma }
\label{eqn:Jbar_ACBN0}
\end{align}
\end{subequations}
respectively, where the summations over $m$, $m'$ $m''$, and $m'''$ are to be understood as restricted to $m, m', m'',m''' \in (\calI,A)$. In Eqs.~\eqref{eqn:Ubar_ACBN0} and \eqref{eqn:Jbar_ACBN0}, $\langle \phi^{(\DFTplusUgroup,A)}_{m'} \phi^{(\DFTplusUgroup,A)}_{m'''} || \phi^{(\DFTplusUgroup,A)}_{m}  \phi^{(\DFTplusUgroup,A)}_{m''}  \rangle$ represents the Coulomb integral in the chemists' notation, $N_{m}^{\sigma}$ represents the formal occupation number of the localized orbital $m$.
In close analogy to Dudarev's formulation, we can evaluate these denominators in Eqs.~\eqref{eqn:Ubar_ACBN0} and~\eqref{eqn:Jbar_ACBN0} as
\begin{subequations}
\begin{align}
    \sum_m \left(N^{(\DFTplusUgroup,A)}\right)_m^\sigma = \Tr{\rho^{(\DFTplusUgroup,A)}_\sigma}
    \label{eqn:trN}
\end{align}
and
\begin{align}
    \sum_{m\neq m'} \left(N^{(\DFTplusUgroup,A)}\right)_m^\sigma \left(N^{(\DFTplusUgroup,A)}\right)_{m'}^\sigma=
    \left( \Tr{\rho^{(\DFTplusUgroup,A)}_\sigma} \right)^2 - 
    \Tr{ \rho^{(\DFTplusUgroup,A)}_\sigma \rho^{(\DFTplusUgroup,A)}_\sigma},
    \label{eqn:trNN}
\end{align}
\label{eqn:trNsigma}
\end{subequations}
respectively.

Traces appearing in Eqs.~\eqref{eqn:trN}~and~\eqref{eqn:trNN} are evaluated as $\Tr{\rho^{(\DFTplusUgroup,A)}_\sigma}= \Tr{Q^{(\DFTplusUgroup,A)}_\sigma}$ and $ \Tr{ \rho^{(\DFTplusUgroup,A)}_\sigma \rho^{(\DFTplusUgroup,A)}_\sigma}=\Tr{ Q^{(\DFTplusUgroup,A)}_\sigma Q^{(\DFTplusUgroup,A)}_\sigma}$, respectively, where the matrix $Q^{(\DFTplusUgroup,A)}_\sigma$ is defined as
\begin{align}
  \left( Q^{(\DFTplusUgroup,A)}_\sigma\right)_{mm'} &= \sum_{\bfk} \sum_{\mu,\nu}\langle \phi^{(\DFTplusUgroup,A)}_m| \chi_{\mu\bfk}\rangle D^{\mu\nu}_{\bfk\sigma}\langle \chi_{\nu\bfk}|\phi^{(\DFTplusUgroup,A)}_{m'} \rangle \nonumber\\
  &= \frac{1}{N_L} \sum_{\bfk} \sum_{n',n'',\mu,\nu}c^{(\calI)}_{n'} S_\bfk^{(n'A\ell_\calI m)\mu} D^{\mu\nu}_{\bfk\sigma} S_\bfk^{\nu(n''A\ell_\calI m')}c^{(\calI)}_{n''}
\end{align}

\subsubsection{Linear response methods}\label{subsubsec:LinearResponseMethods}
Linear response methods are based on the response of the system to an externally applied perturbation on the projected space. 
Cococcioni~\textit{et al.}~\cite{Cococcioni_DFT+Ulr_PRB2005} developed a scheme based on the density response $\chi$ and its Kohn-Sham correspondence $\chi_{KS}$. Their theory has been successfully applied to \textit{ab initio} calculations of correlated systems~\cite{Cococcioni_DFT+Ulr_Review}, including phonon spectrum calculations~\cite{Gross-Cococcioni_phonon_spectrum}, applications to transition-metal chemistry~\cite{Marzari-Cococcioni_TMchemistry}, etc.
Later, Moynihan~\textit{et al.}~\cite{Moynihan-ORegan2017self-consistentU} developed the minimum-tracking linear-response method~\cite{Linscott-ORegan_MTLR_PRB2018}, which is a new formulation of the linear-response method based on the response
of the averaged effective electron-electron interaction $V_{Hxc}$ in the projected subspace.
In this paper, we work on the latter formulation, since it has a computational advantage that the Hubbard parameters can be computed from the site-diagonal responses (i.e., the Hubbard parameters are calculated from the response of the perturbed atom). 


The minimum-tracking linear-response method~\cite{Moynihan-ORegan2017self-consistentU,Linscott-ORegan_MTLR_PRB2018} was first implemented in CP2K by Chai~\textit{et al.}~\cite{Ziwei} in the $\Gamma$-point formulation. In this work, we extend it to $\bfk$-point sampling calculations.
We follow Moynihan~\textit{et al.}~\cite{Moynihan-ORegan2017self-consistentU} and Chai~\textit{et al.}~\cite{Ziwei} for its formulation.
In the minimum-tracking linear-response method, we apply a small perturbation on a specific set of orbitals, $(\calI,A)$, 
\begin{align}
    V^{\mathrm{Pert.}}_\sigma \equiv \lambda^{(\DFTplusUgroup,A)}_\sigma \calP^{(\DFTplusUgroup,A)}
    \label{eqn:Vpert}
\end{align}
with a small parameter $\lambda^{(\DFTplusUgroup,A)}_\sigma$ representing the strength of the perturbation. In periodic solids, the perturbation is to be understood as applied to all unit cells so that $V^{\mathrm{Pert.}}_\sigma$ is lattice periodic. We then calculate the system's response in the Hartree-and-xc (Hxc) potential $V_{\Hxc,\sigma}$ by its orbital average
\begin{align}
    \overline{V}^{(\DFTplusUgroup,A)}_{\Hxc,\sigma}\equiv 
    \frac{1}{N_o}\sum_m \langle \phi_m^{(\calI,A)}|V^{\Hxc}_\sigma |\phi_m^{(\calI,A)}\rangle,
\end{align}
with $N_o$ being the number of orbitals in the set $\{\phi_m^{(\calI,A)}\}$,
and the total occupation number of the local orbitals
\begin{align}
    N_\sigma^{(\DFTplusUgroup,A)}\equiv \Tr{\calP^{(\DFTplusUgroup,A)}\rho_\sigma \calP^{(\DFTplusUgroup,A)}}
\end{align}
to get the $\Ubar$ and $\Jbar$ by the ansatz
\begin{subequations}
\begin{align}
    \Ubar^{(\DFTplusUgroup,A)}= \frac{1}{2} \frac{\partial  \left( \overline{V}_{\Hxc\;\uparrow}^{(\calI,A)} + \overline{V}_{\Hxc\;\downarrow}^{(\calI,A)} \right) }
                     {\partial  \left( N_{\uparrow}^{(\calI,A)} + N_{\downarrow}^{(\calI,A)} \right) } 
    \label{eqn:Ubar_MTLR}
\end{align}
\begin{align}
    \Jbar^{(\DFTplusUgroup,A)}= -\frac{1}{2} \frac{\partial  \left( \overline{V}_{\Hxc\;\uparrow}^{(\calI,A)} - \overline{V}_{\Hxc\;\downarrow}^{(\calI,A)} \right) }
                     {\partial  \left( N_{\uparrow}^{(\calI,A)} - N_{\downarrow}^{(\calI,A)} \right) }
    \label{eqn:Jbar_MTLR}
\end{align}
\end{subequations}
The expression of the perturbation operator $V^{\mathrm{Pert.}}_\sigma$ in the AO representation is given as
\begin{align}
    \langle \chi_{\mu\bfk}| V^{\mathrm{Pert.}}_\sigma |\chi_{\nu\bfk} \rangle = \lambda^{(\DFTplusUgroup,A)}_\sigma \sum_{m,n,n'} \left( S_{\bfk} \right)_{\mu (A n' \ell_\calI m)}c^{(\calI)}_{n'}
c^{(\calI)}_{n} \left( S_{\bfk} \right)_{(A n \ell_\calI m)\nu},
\end{align}
and that of $\overline{V}^{(\DFTplusUgroup,A)}_{\Hxc,\sigma}$ is given as
\begin{align}
    \overline{V}^{(\DFTplusUgroup,A)}_{\Hxc,\sigma} = \frac{1}{N_L} \sum_{\bfk} \sum_{m,n,n'}  c^{(\calI)}_{n'} \left( V_{\bfk \sigma}^{\Hxc} \right)_{(An'\ell_\calI m),(An\ell_\calI m)} c^{(\calI)}_n, 
\end{align}
whereas that of $N_\sigma^{(\DFTplusUgroup,A)}$ is the same as that of Eq.~\eqref{eqn:trN} discussed in subsec.~\ref{subsubsec:ACBN0}

In this study, we parametrize the perturbation strength $\lambda^{(\DFTplusUgroup,A)}_\sigma$ 
in Eq.~\eqref{eqn:Vpert} 
by the charge ($\Delta_c$) and spin ($\Delta_s$) components as
\begin{align}
    \lambda^{(\DFTplusUgroup,A)}_\sigma = \Delta_c +  \Ifrac{\sigma\Delta_s}{2},
\label{eqn:lambda_ChargeAndSpin}
\end{align}
where $\sigma$ represents the sign of the spin projection, which takes the value either $1$ or $-1$. 
When we calculate $\overline{U}^{(\DFTplusUgroup,A)}$ [ $\overline{J}^{(\DFTplusUgroup,A)}$ ], we set only the charge [spin] component finite while the other one is set to $0$.

\subsubsection{Comparison between methodologies}
The Hubbard parameters $U$ and $J$ are characterized as the screened Coulomb and exchange interactions averaged over the orbitals in the projected space, respectively.
In the linear response approaches, the screening is explicitly calculated from the linear response, while in ACBN0, the static correlation among KS orbitals is included in the renormalized density matrix.

A remarkable point in ACBN0 is that, since $U$ and $J$ are calculated only from the static correlations included in the density matrix, it directly applies to real-time simulations in the adiabatic scheme to get time-dependent Hubbard parameters. Such applications were pioneered by Tancogne-Dejean et al.~\cite{Rubio_tdDFT_ACBN0_2017} with a number of applications~\cite{Rubio_MoTe2_LifshitzTransition,Tancogne-Dejean_Rubio_NiO_2020} to time-dependent non-equilibrium dynamics of correlated materials. 
However, we find that the idea of the \textit{renormalized density matrix} and/or the expression of the Hubbard parameters $\Ubar$ and $\Jbar$ (Eqs.~\eqref{eqn:Ubar_ACBN0} and ~\eqref{eqn:Jbar_ACBN0}) are not derived from the standard many-body theory.
This makes it difficult to analyze or predict the behaviors of Hubbard parameters computed using ACBN0. 
Such difficulty manifests in dynamical applications, where it is expected that electronic excitations dynamically renormalize the Hubbard parameters~\cite{Golez-Eckstein-Werner_GW+EDMFT_BandRenormalization,Tancogne-Dejean_Rubio_NiO_2020,Rubio_attosecond_response_RT-TDDFT} by enhanced screening. 
Although the numerical results shown in Ref.~\cite{Rubio_attosecond_response_RT-TDDFT} show striking agreement with experimental results, to the authors' knowledge, there is no 
theory that shows how Eqs.~\eqref{eqn:Ubar_ACBN0} and~\eqref{eqn:Jbar_ACBN0} reproduce such screening effects, how the \textit{renormalized} density matrix (Eq.~\eqref{eqn:renormalizedDM_AORep}) behaves in the dynamics and what limitations exist in this approach.

On the other hand, the linear response theories are derived as screened interactions in the framework of standard many-body theory. They can therefore be extended to dynamical responses, as we show in Sec.~\ref{subsec:Energy-dependent-U}.

\subsection{Energy-dependent Hubbard parameters}\label{subsec:Energy-dependent-U}
The Hubbard parameters are dependent on the energy of quasiparticles due to the energy dependence of screening. 
The effect of such energy dependence of the screened interaction on the self-energy was analyzed in detail by Aryasetiawan~\textit{et al.} in Ref.~\cite{Aryasetiawan-Imada2004_HubbardU_cRPA}.
Their analyses showed a relevant contribution of the high-energy part to the self-energy, which has to be appropriately renormalized to get a low-energy single band model. 
They integrated the high-energy part using the energy-dependent screened interaction calculated using cRPA~\cite{Aryasetiawan-Imada2004_HubbardU_cRPA}.
More recently, Vanzini~\textit{et al.}~\cite{Marzari_DFT+Uw} proposed DFT+$U(\omega)$, which is an extension of DFT+U to incorporate energy-dependent $U$. 
It was reported in Ref.~\cite{Marzari_DFT+Uw_BiFeO3_arXiv2025} that, in the calculation of multiferroic BiFeO$_3$~\cite{Marzari_DFT+Uw_BiFeO3_arXiv2025}, their new scheme showed a qualitative improvement over DFT+U or 
its extension with intersite interactions, DFT+U+V~\cite{DFT+U+V} with static Hubbard parameters.

The most established approach of obtaining $U(\omega)$ is the constrained RPA~\cite{Aryasetiawan-Imada2004_HubbardU_cRPA} (cRPA). In cRPA, the screened interaction is approximated as
\begin{align}
\hat{W}(\omega) = \frac{1}{1-\hat{\Pi}^{(\mathsf{C})}_r(\omega) \hat{v}_{\mathrm{Coul.}} }\hat{v}_{\mathrm{Coul.}}
\label{eqn:W_cRPA}
\end{align}
with $\hat{v}_{\mathrm{Coul.}}$ being the Coulomb interaction operator, $\Pi^{(\mathsf{C})}_r(\omega)$ being the constrained polarization function in the causal ($(\mathsf{C})$) form\cite{RESPACK_cRPA}
\begin{align}
    \Pi^{(\mathsf{C})}_r(\bfr,\bfr';\omega) \equiv \sum_\sigma \sum_{\substack{\bfk,a \in uocc.\\ \bfk',i \in occ.}} &\left(\frac{
     \psi^{\ast}_{a \bfk\sigma}(\bfr)\psi_{i\bfk'\sigma}(\bfr) 
     \psi^{\ast}_{i\bfk'\sigma}(\bfr')\psi_{a\bfk \sigma}(\bfr')
     }{ \hbar\omega - (\varepsilon_{a\bfk \sigma} - \varepsilon_{i\bfk' \sigma}) + i\eta }
     - \frac{
     \psi^{\ast}_{i \bfk' \sigma}(\bfr)\psi_{a \bfk \sigma}(\bfr) 
     \psi^{\ast}_{a \bfk \sigma}(\bfr')\psi_{i \bfk' \sigma}(\bfr')
     }{ \hbar\omega + (\varepsilon_{a \bfk \sigma} - \varepsilon_{i \bfk' \sigma}) - i\eta } \right) \nonumber\\ &\times\left( 1 - \langle a \bfk | \calP | a \bfk \rangle\langle i \bfk' | \calP | i \bfk' \rangle \right),
     \label{eqn:Pi_cRPA}
\end{align}
where $\eta$ is an infinitesimally small positive constant, $\calP$ represents the projection to the correlated orbital space.
The Hubbard parameter $U(\omega)$ is then obtained as an orbital average of the screened interaction $\hat{W}(\omega)$.
Equation~\eqref{eqn:W_cRPA} compactly represents, within the RPA, the Coulomb interaction screened by electrons in \textit{screening orbitals}, which are orbitals other than the correlated orbitals. It can also be easily shown that, within the RPA, $\hat{W}(\omega)$ screened by the electrons in the correlated orbitals equals fully screened Coulomb interaction~\cite{Aryasetiawan-Imada2004_HubbardU_cRPA}. 
cRPA has been established as one of the most standard approaches for the calculation of energy-dependent Hubbard parameters with successful applications~\cite{Miyake-Aryasetiawan-Imada_cRPA}.
On the other hand, there are also known problems in cRPA. 
Honerkamp~\textit{et al.}~\cite{Werner_limitations_cRPA_PhysRevB.98.235151}, using the functional renormalization group analysis, showed that the higher-order diagrams missing in RPA largely cancel the RPA contributions, indicating that cRPA overestimates screening. 
%
%
Carta~\textit{et al.}~\cite{Carta-Timrov_Bridging_cRPA_and_LR_2505.03698v1}, who worked on the calculation of static Hubbard parameters using cRPA and Cococcioni's linear-response theory~\cite{Cococcioni_DFT+Ulr_PRB2005}, pointed out the problem arising from the distinction between correlated and screening orbitals. They reported overscreening in cRPA for materials with strong hybridization among two types of orbitals. They also reported robustness of the linear response theory even in those difficult cases.
We can also see that the screening in Eqs.~\eqref{eqn:W_cRPA} and ~\eqref{eqn:Pi_cRPA} 
is 
derived with only the direct Coulomb interactions taken into account. 
This is, in general, not desirable for the calculation of Hubbard parameters to be put into DFT+U calculations, since we expect that the appropriate values of $U$ and $J$, such that they approximately remove the self-interaction error in DFT, should be dependent on the choice of xc functional.

\subsubsection{Extension of linear response schemes}\label{subsubsec:Extension_of_LR_schemes}
The fact that cRPA is not a perfect tool for calculating $U(\omega)$ motivates us to develop a possible alternative way to calculate $U(\omega)$ using linear response theories.
We first rewrite Eq.~\eqref{eqn:Ubar_MTLR} as follows;
\begin{align}
    \Ubar^{(\DFTplusUgroup,A)} &= \frac{1}{2} \frac{1}{N_o} \sum_m 
    \left \langle \phi_m^{(\DFTplusUgroup,A)} \left| 
    \frac{\partial  \left( V^{\Hxc}_{\uparrow} + V^{\Hxc}_{\downarrow} \right) }{\partial  \left( N_{\uparrow}^{(\calI,A)} + N_{\downarrow}^{(\calI,A)} \right) } \right|  \phi_m^{(\DFTplusUgroup,A)} \right\rangle \nonumber\\
    &= \frac{1}{2} \frac{1}{N^2_o} \sum_{m,m'} \left \langle \phi_m^{(\DFTplusUgroup,A)} \left| 
    \frac{\partial  \left( V^{\Hxc}_{\uparrow} + V^{\Hxc}_{\downarrow} \right) }{\partial  \left( n_{m'\uparrow}^{(\calI,A)} + n_{m'\downarrow}^{(\calI,A)} \right) } \right|  \phi_m^{(\DFTplusUgroup,A)} \right\rangle \nonumber\\
    &= \frac{1}{4} \sum_{\substack{m,m'\\\sigma,\kappa}}\frac{1}{N^2_o}
    \left \langle \phi_m^{(\DFTplusUgroup,A)} \left| 
    \frac{\partial V^{\Hxc}_{\sigma}}{\partial  n_{m'\kappa}}  
    \right|  \phi_m^{(\DFTplusUgroup,A)} \right\rangle
    \nonumber\\&
    = \frac{1}{4} \sum_{\substack{m,m'\\\sigma,\kappa,\upsilon}}\frac{1}{N^2_o} \iint d^3\bfr d^3\bfr' \left|\phi_m^{(\DFTplusUgroup,A)}(\bfr) \right|^2 
    \frac{ \delta V_\sigma^{\Hxc}(\bfr) }{ \delta \density_\upsilon (\bfr')}
    \frac{ \partial \density_\upsilon(\bfr')}{\partial n_{m'\kappa}}
    \nonumber\\&
    = \frac{1}{4} \sum_{\substack{m,m'\\\sigma,\kappa,\upsilon}}\frac{1}{N^2_o} \iint d^3\bfr d^3\bfr' \left|\phi_m^{(\DFTplusUgroup,A)}(\bfr) \right|^2 
    f_\Hxc^{\sigma\upsilon}(\bfr,\bfr')
    \frac{ \partial \density_\upsilon(\bfr')}{\partial n_{m\kappa}}
    \label{eqn:Ubar_response1}
\end{align}
with $\sigma,\kappa,\upsilon$ representing the spin projection and $f^{\sigma\upsilon}_{\Hxc}(\bfr,\bfr') \equiv \Ifrac{\delta V^{\Hxc}_{\sigma}(\bfr)}{\delta \density_{\upsilon}(\bfr')}$ representing the Hartree-xc kernel. In the second line of Eq.~\eqref{eqn:Ubar_response1} we replaced the derivative $\frac{\partial}{\partial \left( N_{\uparrow}^{(\calI,A)} + N_{\downarrow}^{(\calI,A)} \right)}$ by $\frac{1}{N_o}\sum_{m'} \frac{\partial}{\partial (n_{m'\uparrow}^{(\calI,A)} + n_{m'\downarrow}^{(\calI,A)}) } = \frac{1}{2N_o}\sum_{m',\kappa} \frac{\partial}{\partial n_{m'\kappa}} $.
The density variation $\delta \density_\upsilon(\bfr')$ induced by an infinitesimal variation of the occupation number $\delta n_{m'\kappa}$ is written as the sum of the 'direct' contribution and that induced by the variation of the Hxc potential as
\begin{align}
    \delta \density_\upsilon(\bfr')= \int d^3\bfr'' \left( \delta_{\upsilon\kappa}\delta^3(\bfr'-\bfr'') + \sum_{\kappa'}\int d^3\bfr''' \chi_{\upsilon\kappa'}(\bfr',\bfr''')f^{\kappa'\kappa}_{\Hxc}(\bfr''',\bfr'') \right) |\phi^{(\DFTplusUgroup,A)}_{m'}(\bfr'')|^2 \delta n_{m'\kappa}
    \label{eqn:deltarho_LR_static}
\end{align}
with $\chi_{\upsilon\kappa'}(\bfr',\bfr'')\equiv \Ifrac{\delta \density_{\upsilon}(\bfr')}{\delta V^{\ext}_{\kappa'}(\bfr'')}$ being the static (spin-)density response function to an infinitesimal external perturbation $\delta V^{\ext}_{\kappa'}(\bfr'')$ applied to electrons with spin projection $\kappa'$. Substituting Eq.~\eqref{eqn:deltarho_LR_static} into Eq.~\eqref{eqn:Ubar_response1}, we get
{\small
\begin{align}
    \Ubar^{(\DFTplusUgroup,A)} &= \frac{1}{4} \sum_{\sigma,\kappa,\upsilon} \frac{1}{N_o^2}\sum_{m,m'} \iiint d^3\bfr d^3\bfr'd^3\bfr''
    |\phi^{(\DFTplusUgroup,A)}_{m}(\bfr)|^2
    f^{\sigma\upsilon}_{\Hxc}(\bfr,\bfr') 
    \nonumber\\&\hspace{10pt}\times
    \left( \delta_{\upsilon\kappa}\delta^3(\bfr'-\bfr'')
    + \sum_{\kappa'}\int d^3\bfr''' \chi_{\upsilon,\kappa'}(\bfr',\bfr'') f^{\kappa'\kappa}_{Hxc}(\bfr''',\bfr'') \right) | \phi^{(\DFTplusUgroup,A)}_{m'}(\bfr'')|^2 \nonumber\\
    &= \frac{1}{4} \sum_{\sigma,\kappa,\upsilon} \frac{1}{N_o^2}\sum_{m,m'} \iiint d^3\bfr d^3\bfr' d^3\bfr''
    |\phi^{(\DFTplusUgroup,A)}_{m}(\bfr)|^2 
    f^{\sigma\upsilon}_{\Hxc}(\bfr,\bfr') 
    \epsilon^{-1}_{\upsilon,\kappa}(\bfr',\bfr'')| \phi^{(\DFTplusUgroup,A)}_{m'}(\bfr'')|^2
\end{align}
}
where we introduced the screening function $\epsilon^{-1}$, which is symbolically defined as $\epsilon^{-1} = 1 + \chi f_{\Hxc}$ (we note that this definition is analogous to the \textit{exact} expression in many-body theory~\cite{Martin_Book} $\epsilon^{-1} = 1 + \chi v_{\mathrm{Coul.}}$). 
We can extend this discussion in a time-dependent manner as
\begin{align}
    &\delta V^{\Hxc}_\sigma(\bfr,t) = 
    \int^t dt' \sum_\upsilon\iint d^3\bfr' d^3\bfr'' f_{\Hxc}^{\sigma\upsilon}(\bfr,\bfr')
    \nonumber\\&\hspace{10pt}\times
    \left(
      \delta_{\sigma\upsilon}\delta^3(\bfr'-\bfr'')\delta(t-t')
      + \sum_{\kappa'}\int d^3\bfr''' \chi^{\upsilon\kappa'}(\bfr',\bfr''';t-t') f^{\kappa'\kappa}_{\Hxc}(\bfr''',\bfr'') \right) 
      | \phi^{(\DFTplusUgroup,A)}_{m'}(\bfr'')|^2 \delta n_{m'}^{(\DFTplusUgroup,A)}(t')
\end{align}
where we assumed the adiabaticity of the xc functional (i.e., we assumed that the Hxc kernel $f_\Hxc$ is an instantaneous function of time). We therefore get
\begin{subequations}
\begin{align}
    \delta \overline{U}^{(\DFTplusUgroup,A)}(t)
    = \int^t dt' \frac{1}{4 N_o^2} \sum_{m,m',\sigma,\kappa} 
      \calU^{\sigma\kappa}_{m,m'}(t-t') \delta n_{m'\kappa}(t') 
      \label{eqn:tdResponse_MTLR_U}
\end{align}
\begin{align}
    \delta \overline{J}^{(\DFTplusUgroup,A)}(t)
    = \int^t dt' \frac{-1}{4 N_o^2} \sum_{m,m',\sigma,\kappa} 
      \sigma\kappa \calU^{\sigma\kappa}_{m,m'}(t-t') \delta n_{m'\kappa}(t') 
    \label{eqn:tdResponse_MTLR_J}
\end{align}
\end{subequations}
with $\calU^{\sigma\kappa}_{m,m'}(t-t')$ being the response kernel
\begin{align}
    \calU^{\sigma\kappa}_{m,m'}(t-t') &\equiv 
    \sum_\upsilon\iiint d^3\bfr d^3\bfr' d^3\bfr'' 
    |\phi_{m}^{(\DFTplusUgroup,A)}(\bfr)|^2
    f_{\Hxc}^{\sigma\upsilon}(\bfr,\bfr')
    \nonumber\\&\hspace{10pt}\times
    \left(
      \delta_{\sigma\upsilon}\delta^3(\bfr'-\bfr'')\delta(t-t')
      + \sum_{\kappa'}\int d^3\bfr''' \chi^{\upsilon\kappa'}(\bfr',\bfr''';t-t') f^{\kappa'\kappa}_{\Hxc}(\bfr''',\bfr'') \right) 
      | \phi^{(\DFTplusUgroup,A)}_{m'}(\bfr'')|^2.
      \label{eqn:Umm_time}
\end{align}
We can apply the linear response technique in RT-TDDFT~\cite{Yabana-Bertsch1996} to get the response kernels in the frequency domain in the retarded ($(\mathsf{R})$) form, which are, for Eq.~\eqref{eqn:tdResponse_MTLR_U},
\begin{align}
    \delta \overline{U}^{(\DFTplusUgroup,A)}(\omega) = 
    \frac{1}{4 N_o^2} \sum_{m,m',\sigma,\kappa} 
    \calU^{\sigma\kappa\;(\mathsf{R})}_{m,m'}(\omega) \delta n_{m'\kappa}(\omega),
    \label{eqn:Ubar_response}
\end{align}
where $\overline{U}^{(\DFTplusUgroup,A)}(\omega)$ is the Fourier transformation of Eq.~\eqref{eqn:Umm_time}. For the sake of clarity, we rewrite it in a symbolic form as
\begin{align}
   \calU^{(\mathsf{R})}_{m,m'}(\omega) = \langle  \phi_{m}^{(\DFTplusUgroup,A)} \phi_{m}^{(\DFTplusUgroup,A)} | \hat{f}_{\Hxc} \left( \underline{1} + \chi^{(\mathsf{R})}(\omega)\hat{f}_{Hxc} \right) | \phi_{m'}^{(\DFTplusUgroup,A)} \phi_{m'}^{(\DFTplusUgroup,A)} \rangle,
   \label{eqn:Umm_omega}
\end{align}
where $\calU^{(\mathsf{R})}_{m,m'}$, $\hat{f}_{\Hxc}$, and $\chi^{(\mathsf{R})}(\omega)$ without Greek (spin) superscripts are to be understood as $2\times 2$ matrices, and $\underline{1}$ is to be understood as the $2\times 2$ unit matrix.
Here $\calU^{\sigma\kappa\;(\mathsf{R})}_{m,m'}(\omega)$ represents the retarded form of the frequency-domain response function $\calU^{\sigma\kappa\;(\mathsf{R})}_{m,m'}(\omega)=\int_0^{\infty} dt e^{i\omega t} \calU^{\sigma\kappa}_{m,m'}(t)$,
which contains the retarded form of the screening function $\epsilon^{-1\;(\mathsf{R})} = \underline{1} + \chi^{(\mathsf{R})}f_{Hxc}$. 
The analytic property of the retarded response function is different from what is obtained using cRPA, in which the screening function is derived in the causal form; however, the real part of the response for positive frequency $\omega>0$, which is of our interest, should be the same for causal and retarded form. 

We follow the standard procedure in RT-TDDFT linear response calculations. 
We apply the impulsive form of perturbation $V^{\mathrm{Pert.}}_\sigma(t) = \lambda^{(\DFTplusUgroup,A)}_\sigma \delta(t) \calP^{(\DFTplusUgroup,A)}$ at $t=0$ and collect the response of $N_{\uparrow}^{(\calI,A)} \pm N_{\downarrow}^{(\calI,A)}$ and $\overline{V}_{\Hxc\uparrow}^{(\calI,A)}\pm \overline{V}_{\Hxc\downarrow}^{(\calI,A)}$ in the form of a time series along the RT-TDDFT simulation from $t=0$ to $t=T$.
We then Fourier-transform each observable to get
\begin{subequations}
\begin{align}
    \Ubar^{(\DFTplusUgroup,A)}(\omega)\approx \frac{1}{2} \frac{\delta  \left( \overline{V}_{\Hxc\;\uparrow}^{(\calI,A)}(\omega) + \overline{V}_{\Hxc\;\downarrow}^{(\calI,A)}(\omega) \right) }
                     {\delta  \left( N_{\uparrow}^{(\calI,A)} (\omega)+ N_{\downarrow}^{(\calI,A)} (\omega)\right) } 
    \label{eqn:Ubar_tdMTLR}
\end{align}
\begin{align}
    \Jbar^{(\DFTplusUgroup,A)}(\omega)\approx -\frac{1}{2} \frac{\delta  \left( \overline{V}_{\Hxc\;\uparrow}^{(\calI,A)}(\omega) - \overline{V}_{\Hxc\;\downarrow}^{(\calI,A)}(\omega) \right) }
                     {\delta  \left( N_{\uparrow}^{(\calI,A)}(\omega) - N_{\downarrow}^{(\calI,A)}(\omega) \right) }.
    \label{eqn:Jbar_tdMTLR}
\end{align}
\end{subequations}
Although Eq.~\eqref{eqn:Ubar_tdMTLR} is different from the microscopic expression Eq.~\eqref{eqn:Ubar_response} 
(the occupation numbers in Eq.~\eqref{eqn:Ubar_tdMTLR} are summed over the orbital indices),
Eq.~\eqref{eqn:Ubar_tdMTLR} works as a direct extension of the static expression, i.e.,
$\Ubar^{(\DFTplusUgroup,A)}$ calculated with Eq.~\eqref{eqn:Ubar_MTLR} equals
$\Ubar^{(\DFTplusUgroup,A)}(\omega)$ calculated with Eq.~\eqref{eqn:Ubar_tdMTLR} at $\omega=0$.    

We also note that both $\Ubar^{(\DFTplusUgroup,A)}(\omega)$ and $\Jbar^{(\DFTplusUgroup,A)}(\omega)$ are complex-valued quantities. Their real parts work as the effective potential for quasiparticles of energy $\hbar\omega$, whereas their imaginary parts arise from poles of the density correlation function (see Eq.~\eqref{eqn:Umm_omega}). Details are discussed in Sec.~\ref{subsec:RSL_MTLR_dynamical}.  

To get time-dependent dynamical responses for the calculation of $\overline{U}^{(\calI,A)}(\omega)$ [$\overline{J}^{(\calI,A)}(\omega)$], we set the charge [spin] component of the impulse strength to a positive small value $\Delta$ as $\Delta_c = \Delta $
[ $\Delta_s = \Delta$ ] in simulation $A$ and 
$\Delta_c = -\Delta $ [ $\Delta_s = -\Delta$ ] in simulation $B$. 
The dynamical responses are derived as the difference between observables from simulations $A$ and $B$. 

In principle, we should also be able to extend Cococcioni's linear response theory~\cite{Cococcioni_DFT+Ulr_PRB2005} (see also Appendix~\ref{app:CococcioniLRT} for our notations) to an energy-dependent form as
\begin{align}
    \Ubar^{I}(\omega)=\left( \chi_{\KS}^{-1}(\omega) - \chi^{-1}(\omega) \right)^{II},
    \label{eqn:Uw_in_dLR}
\end{align}
as a direct extension of the static expression (Eq.~\eqref{eqn:U_in_dLR} in Appendix~\ref{app:CococcioniLRT}). To compute quantities in Eq.~\eqref{eqn:Uw_in_dLR}, one can, in principle, calculate the dynamical response of the system to an impulsive perturbation to get $\chi(\omega)$ (as a retarded response function), whereas the energy-dependent $\chi_{\KS}(\omega)$ can be computed analytically. However, we did not explore this possibility because of the computational cost. 
Cococcioni's linear response theory requires calculation with a perturbation on each atom in the supercell. One therefore needs to perform $N_H^{sc}$ numerical differentiation calculations, where $N_H^{sc}$ represents~\cite{Timrov-Marzari-Cococcioni_DFPTlr} the number of atoms with correlated orbitals in the supercell. 
In our dynamical linear response scheme, this means that, to get $U(\omega)$, one would have to calculate $2N_H^{sc}$ trajectories (factor $2$ for $\Delta_c=\Delta$ and $\Delta_c=-\Delta$), which is prohibitively expensive even for moderately large supercells. 

\section{Calculation results} \label{sec:CalculationResults}
Throughout this paper, numerical calculations are performed using CP2K in the pseudopotential formulation using the Goedecker-Teter-Hutter (GTH) pseudopotential~\cite{GTHpsp}. The PBE~\cite{PBE} xc functional, and TZVP-MOLOPT~\cite{MOLOPTbset} basis set are used, unless stated otherwise.
We show results of static and dynamical calculations using ACBN0 and the minimum-tracking linear-response method. Some results of Cococcioni's linear response theory are shown in Appendix~\ref{app:CococcioniLRT}.

\subsection{ACBN0, static calculations}\label{subsec:RSL_ACBN0_static}
To check the validity of our implementation of ACBN0, we first calculated static properties of $6$ transition metal oxides/nitrides 
(TMO/Ns), which are ScN, Rutile TiO$_2$, MnO, NiO, Wurtzite ZnO, and CuO.
In this paper, TiO$_2$ and ZnO are assumed to be of the Rutile structure and Wurtzite structure, respectively. 
The unit cell configurations of these materials are shown in the Supporting Material.
We applied the on-site repulsion $U_{\eff}=U-J$ on both the transition metal (TM) $3d$-band and the oxygen or nitrogen $2p$-band.
The Brillouin zone was sampled using a $16\times 16\times 16$ Monkhorst-Pack~\cite{Monkhorst-Pack} mesh. 
We first show the obtained values of $U_{\eff}$, band gaps, and magnetic moments in Table~\ref{table:ACBN0}.
\begin{table}[ht]
\begin{tabular}{|c|c|c|c|c|c|c|}
\hline
 & \multicolumn{4}{|c|}{calculation results} & \multicolumn{2}{|c|}{experimental results} \\
     & U$_{\eff}$ (M) & U$_{\eff} $ (O/N) & Band Gap {\small (indirect/direct)} & m [$\mu_B$] & Band Gap & m [$\mu_B$] \\
\hline
ScN	    & 1.679	& 6.539	& 1.103/2.321  &	& $0.9 \pm 0.1/2.15$\cite{ScN_BG} & \\	
\hline
TiO$_2$ & 0.918	& 8.987	& 3.101		   &    & $3.3 \pm 0.5$\cite{TiO2_BG} &	     \\
\hline
MnO	    & 4.255	& 4.704	& 2.682/3.163  & 4.590	& 4.1\cite{MnO_BG}	& 4.58\cite{MnO-NiO_magmom} \\
\hline
NiO	   & 3.605	& 3.653	& 3.119/3.490  & 1.560	& 4.2\cite{NiO_BG}	& 1.9\cite{MnO-NiO_magmom} \\
\hline
CuO	   & 3.690	& 2.768	& 2.671	       & 1.584 &	$1.4\pm0.3$\cite{CuO_BG}/$1.6777$\cite{CuO_optBG} & 0.38\cite{CuO_magmom} \\
\hline
ZnO	   & 14.255	& 5.997	& 3.336	       &        & 3.4\cite{ZnO_BG}	&      \\ 
\hline
\end{tabular}
\caption{Results of static ACBN0 calculations. The first $4$ columns show the calculated values of Hubbard parameters $U_{\eff}$ in eV for $3d$ orbitals on the metallic (M) atoms and $2p$ orbitals on the oxygen/nitride (O/N) atoms, the band gap in eV (indirect/direct if they are different), magnetic moment in the scale of the Bohr magneton $\mu_B$. The 5th and 6th columns show the corresponding experimental results.     }
\label{table:ACBN0}
\end{table}
The band gaps of ScN, TiO$_2$, ZnO, and the magnetic moment of MnO are in reasonable agreement with the experimental results, while the band gaps of NiO and MnO were underestimated, and that of CuO is overestimated. 
We also compared the obtained values of the Hubbard parameters $U_{\eff}$ with those in the preceding ACBN0 calculation reports by Agapito~\textit{et al}~\cite{Agapito-Nardelli2015ACBN0} and Tancogne-Dejean~\textit{et al.}~\cite{Rubio_tdDFT_ACBN0_2017} in Table~\ref{table:ACBN0_Ueff}, which show large discrepancies.
Since the obtained physical properties (band gaps and magnetic moments) are in reasonable agreement with the experimental results, these discrepancies in the Hubbard parameters are not necessarily problematic.
We attribute the cause of such discrepancies to the difference in the 
pseudopotential~\cite{Marzari_psp_dependence} and the 
basis set, in particular, the basis functions of the projected space, 
among others.
\begin{table}[ht]
\begin{tabular}{|c|c|c|c|c|c|c|}
\hline
 & \multicolumn{3}{|c|}{U$_{\eff}$ (M) (eV)}  
 & \multicolumn{3}{|c|}{U$_{\eff}$ (O) (eV)} \\
 \hline
 & this work & Ref.[\cite{Rubio_tdDFT_ACBN0_2017}] & Ref.[\cite{Agapito-Nardelli2015ACBN0}] 
 & this work & Ref.[\cite{Rubio_tdDFT_ACBN0_2017}] & Ref.[\cite{Agapito-Nardelli2015ACBN0}] \\ 
\hline
TiO$_2$ & 0.918	& 0.96  & 0.15  & 8.987	& 10.18  & 7.34 \\
\hline
MnO     & 4.255	& 4.68  & 4.67  & 4.704	& 5.18  &  2.68 \\
\hline
NiO     & 3.605	& 6.93  & 7.63  & 3.653	& 2.87  &  3.0 \\
\hline
ZnO     & 14.255	& 13.3  & 12.8  & 5.997	& 5.95  &  5.29 \\
\hline
\end{tabular}
\caption{Comparison of the values of the Hubbard parameters with 
those reported in Refs.\cite{Rubio_tdDFT_ACBN0_2017} and \cite{Agapito-Nardelli2015ACBN0}. 
The first 3 columns show $U_\eff$ for metallic 3d orbitals, while the next 3 columns show that for oxygen 2p orbitals.}
\label{table:ACBN0_Ueff}
\end{table}

In these calculations, we applied the on-site repulsion term to both metallic $3d$ orbitals and oxygen/nitride $2p$ orbitals. The magnitudes of these two parameters affect the relative energies of the corresponding (metal $3d$, oxygen/nitride $2p$, and/or their hybrid) bands. We therefore calculated the projected density of states (pDOS) of these materials to check if the obtained band structures reproduce the qualitative nature of the target materials. As shown in Appendix~\ref{app:ACBN0pDOS}, pDOSs of these materials were found 
to be 
qualitatively consistent with known properties.

\subsection{Minimum-tracking linear response, static calculations}\label{subsec:RSL_MTLR_static}
We next show the results of minimum-tracking linear response calculations.
To validate our implementation, we calculated the Hubbard parameters of $4$ types of TMO/N, whose unit cell configurations are shown in the 
Supporting Material.
The perturbation strength $\lambda^{(\DFTplusUgroup,A)}_\sigma$ in Eq.~\eqref{eqn:Vpert} was parameterized by the charge ($\Delta_c$) and spin ($\Delta_s$) components as Eq.~\eqref{eqn:lambda_ChargeAndSpin}.
We applied charge and spin perturbations separately to calculate $\overline{U}^{(\calI,A)}$ (Eq.~\eqref{eqn:Ubar_MTLR}) and $\overline{J}^{(\calI,A)}$ (Eq.~\eqref{eqn:Jbar_MTLR}), respectively.
For each type of perturbation, $\Delta_c$ or $\Delta_s$ was set to $0.05$ eV. Only in the calculations of TiO$_2$, $\Delta_c$ was set to $0.1$ eV. 
We used (non-symmetric) $3$-point differential to numerically evaluate the derivatives in Eq.~\eqref{eqn:Ubar_MTLR} or~\eqref{eqn:Jbar_MTLR}. 
The perturbation was applied to a single atom, and the response of the same atom was measured. The resultant $\overline{U}^{(\calI,A)}$ and $\overline{J}^{(\calI,A)}$ were averaged over the atoms in the unit cell to get the final results, i.e. 
\begin{subequations}
\begin{align}
  \overline{U}^{(\calI)} = \frac{1}{N^{\calI}_A} \sum_A \overline{U}^{(\calI,A)} \\
  \overline{J}^{(\calI)} = \frac{1}{N^{\calI}_A} \sum_A \overline{J}^{(\calI,A)},
\end{align}    
\end{subequations}
with $N_A^{(\calI)}$ being the number of atoms for averaging. 
In the calculations of TiO$_2$, ScN, NiO, and MnO shown below, $N_A^{(\calI)}$ for the metallic [non-metallic] species was set $2$, $4$, $4$, and $4$ [ $4$, $4$, $2$, and $2$ ].
For antiferromagnetic systems, $\overline{U}^{(\calI)}$ and $\overline{J}^{(\calI)}$ for A and B lattices are averaged. We identified the effective Hubbard parameter as $U_\eff^{\calI}=\overline{U}^{(\calI)}-\overline{J}^{(\calI)}$.
In our calculations, we required consistency of the values of the Hubbard parameters used in the calculations (input) and those obtained from linear response analyses (output).  
We iteratively achieved this self-consistency. The Hubbard parameters of ScN, NiO, and MnO were converged to achieve an input-output difference less than $2\times 10^{-6}$ in atomic units (au), whereas those of TiO$_2$ were converged to a difference less than $2.5\times 10^{-4}$ in au.
To reduce numerical errors in large supercell calculations, we applied a symmetry constraint on up and down spins in the antiferromagnetic lattice in the unperturbed ground-state calculations of NiO and MnO, as detailed in Appendix~\ref{app:Enforcing_Symmetry}. 
To get consistent results of the band gap from a limited size of $\bfk$-mesh, we shifted the origin of the Monkhorst-Pack $\bfk$ mesh of even number to include the $\Gamma$-point in the calculations of ScN and MnO.   
We first tested TiO$_2$ $3\times 3\times 5$ supercell with k-mesh set $1\times 1\times 1$ to directly compare our results with the $\Gamma$-point calculation by Chai~\textit{et al.}~\cite{Ziwei}. 
The resultant band gap was 
$\Delta_g = 3.447$ eV, 
which is in good agreement with $\Delta_g =3.44$ eV, reported in Ref.~\cite{Ziwei}.
We next calculated $3$ types of TMO/Ns, 
namely, TiO$_2$, NiO, and ScN, with varying sizes of $\bfk$-point sampling 
to get the results shown in Table~\ref{table:MTLRkp_TMON}.
\begin{table}[ht]
    \centering
    \begin{tabular}{|c|c|c|c|c|c|c|}
    \hline
         & supercell & $\bfk$-mesh & $U_\eff$(M) & $U_\eff$ (O/N) & Band Gap & m [$\mu_B$]\\
\hline
ScN & c$2\times 2\times 2$ & $3\times 3\times 3$  &  2.465 & 7.602 & 1.774 & \\
\hline
    & r$4\times 4\times 4$ & $1\times 1\times 1$  &  2.494 & 7.573 & 1.892 & \\
\hline
    &                      & $2\times 2\times 2$  &  2.499 & 7.548 & 1.908 & \\
\hline
    &                      & $3\times 3\times 3$  &  2.499 & 7.547 & 1.908 & \\
\hline
TiO$_2$ &  $2\times 2\times 3$ & $3\times 3\times 3$ & 3.434 & 9.088 & 3.421 & \\
\hline
        &                      & $4\times 4\times 4$ & 3.433 & 9.112 & 3.529 & \\
\hline
        &  $3\times 3\times 4$ & $1\times 1\times 1$ & 3.475 & 9.175 & 3.440 & \\
\hline
        &                      & $2\times 2\times 2$ & 3.475 & 9.205 & 3.634  & \\
\hline
NiO     &  $2\times 2\times 2$ & $3\times 3\times 3$ & 5.433 & 11.796 & 5.285 & 1.708 \\
\hline
        &                      & $4\times 4\times 4$ & 5.433 & 11.795 & 5.301 & 1.708 \\
\hline
        &  $3\times 3\times 2$ & $1\times 1\times 1$ & 5.562 & 11.913 & 5.690 & 1.707 \\
\hline
        &                      & $2\times 2\times 2$ & 5.532 & 12.027 & 5.500 & 1.712 \\
\hline
        &                      & $3\times 3\times 3$ & 5.532 & 12.027 & 5.381 & 1.712 \\
\hline
MnO     & $2\times 2\times 2$ & $3\times 3\times 3$ & 3.897 & 14.448 & 3.877 & 4.650 \\
\hline
        &  & $4\times 4\times 4$ & 3.894 & 14.432 & 3.875 & 4.650 \\
\hline
        & $3\times 3\times 2$ & $2\times 2\times 2$ & 3.977 & 14.678 & 3.928 & 4.653 \\
\hline
        & & $3\times 3\times 3$ & 3.978 & 14.677 & 3.939 & 4.653 \\
\hline
        & & $4\times 4\times 4$ & 3.978 & 14.677 & 3.928 & 4.653 \\
\hline
    \end{tabular}
    \caption{Results of static minimum-tracking linear response calculations. The first $3$ columns show the material, the supercell size, and the Monkhorst-Pack $\bfk$-mesh. The $4$th to $7$th columns show the corresponding calculated values of $U_{\eff}$ (in eV) for $3d$ orbitals on the metallic (M) atoms and $2p$ orbitals on the oxygen/nitride (O/N) atoms, the (indirect) band gap, magnetic moment in the scale of the Bohr magneton $\mu_B$.
    For ScN, the notation $c 2\times 2\times 2$ means the $2\times 2\times 2$ supercell of the cubic unit cell, while $r 4\times 4\times 4$ indicates the $4 \times 4 \times 4$ supercell of the rhombohedral unit cell.}
    \label{table:MTLRkp_TMON}
\end{table}
From Table~\ref{table:MTLRkp_TMON}, we find that the convergence of the Hubbard parameters with respect to the supercell size is slow, whereas that with respect to the size of $\bfk$-mesh size is fast in the largest supercell of each material shown in Table~\ref{table:MTLRkp_TMON}.
Such dependence on the supercell size 
can
arise from the influence of the perturbation potential $V^{\mathrm{Pert.}}$ (Eq.~\eqref{eqn:Vpert}) acting on the periodic images, hence 
a sufficiently large supercell is required 
even with $\bfk$-point sampling.
The resultant values of the (indirect) band gaps of ScN and NiO are by $\sim1$ eV overestimated compared to the experimental values, whereas those for TiO$_2$ and MnO are in reasonable agreement with experimental values. 
The magnetic moments of NiO and MnO are also in reasonable agreement. 
On the other hand, we find no agreement in the estimated values of the Hubbard parameters between ACBN0 (Table~\ref{table:ACBN0}) and the minimum-tracking linear response (Table~\ref{table:MTLRkp_TMON}).
In terms of the accuracy of these static calculation results, we cannot judge the superiority of either approach since the ScN band gap was better reproduced by ACBN0, while that for MnO was better reproduced by the linear-response method. 
On the other hand, we found that the computational costs were much larger in the minimum-tracking linear response calculations because of the supercell calculations, lack of symmetry, and iterative procedures to get self-consistent values of the Hubbard parameters.

\subsection{ACBN0, dynamical calculations}\label{subsec:RSL_ACBN0_dynamical}
\begin{figure}[ht]
\includegraphics{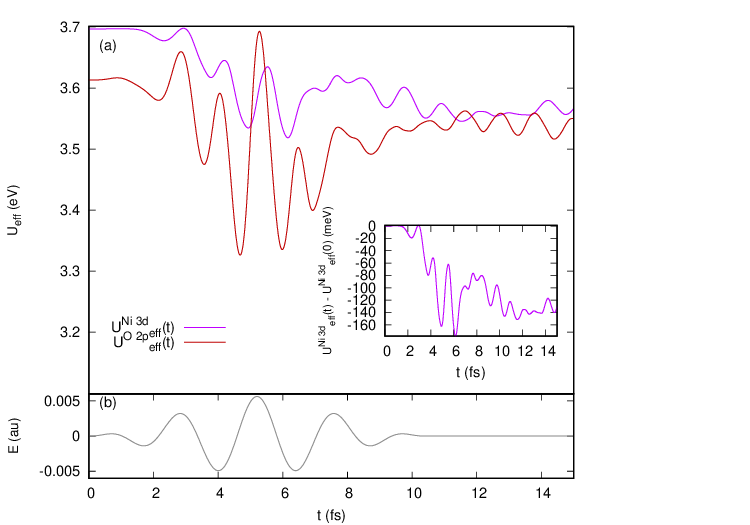}
\caption{Time-dependent Hubbard parameters in the strong-field dynamics. Panel (a) shows the time-dependent behaviors of $U_{\eff}$ for Ni 3d orbitals (purple) and O 2p orbitals (red). The inset shows the difference $U_{\eff}^{\mbox{\scriptsize Ni 3d}}(t)-U_{\eff}^{\mbox{\scriptsize Ni 3d}}(0)$. Panel (b) shows the applied electric field $E(t)=-\frac{1}{c}\frac{d}{dt}A^{\ext}(t)$ as a function of time.}
\label{fig:Ut_ACBN0}
\end{figure}

Non-equilibrium RT-TDDFT+U simulations in a strong laser field using ACBN0 have been pioneered by Tancogne-Dejean \textit{et al.}~\cite{Tancogne-Dejean_Rubio_PRL2018,Tancogne-Dejean_Rubio_NiO_2020}. They applied ACBN0 to calculate time-dependent Hubbard parameters in their simulations of strong-field dynamics of correlated materials. In Ref.~\cite{Tancogne-Dejean_Rubio_NiO_2020}, they found that calculations with time-dependent $U$ reproduce the band-gap renormalization, which represents a dynamical effect in a strongly-correlated system in intense laser fields beyond the dynamical Franz-Keldysh effect. 
Following a recent paper by Cazali~\textit{et al}~\cite{Rubio_attosecond_response_RT-TDDFT}, we simulated the electron dynamics of NiO in a strong laser pulse. 
We modeled the NiO crystal by a 4-atom unit cell of the rock salt structure with lattice constant $4.1704$~\AA. 
The Brillouin zone was sampled using the Monkhorst-Pack scheme with $16\times 16\times 8$ $\bfk$-point mesh.
The external field was modeled as a sine-square pulse of the full-width half maximum 
$5.2$~fs, 
peak intensity $I=1.1052$ TW/cm$^2$. The wavelength was set 
to 
$\lambda = 760$~nm.
Following the formulation developed by Bertsch~\textit{et al.}~\cite{Bertsch-Yabana-Rubio_Aind}, we calculated the internal induced field $\bfA_{\ind}$ by integrating the Maxwell equation $\frac{1}{c^2} \frac{d^2}{dt^2} \bfA_{\ind}(t) = - \frac{4\pi e}{c} \overline{\bfj}(t)$ with $e$ being the elementary charge and $\overline{\bfj}(t)$ being the electron current density averaged over the unit cell.
We set the stepsize $\Delta t=5$~as and propagation time $T=15$~fs.
The resultant time-dependent $U_{\eff}(t)$ are shown in Fig.~\ref{fig:Ut_ACBN0}. 
We find 
that 
the values of $U_{\eff}(t)$ for both Ni 3d and O 2p decrease with the laser excitation. This reduction of $U_{\eff}$ did not disappear even after the pulse duration, indicating that it was caused not directly by the laser field but by the electronic excitation. This behavior is in accord with the results shown in Refs.~\cite{Rubio_attosecond_response_RT-TDDFT} and \cite{Tancogne-Dejean_Rubio_NiO_2020}. 
The amplitude of the reduction $|U(t)-U(t=0)|$ of Ni at the end of the pulse was found to be around $140$ meV, which is slightly larger than the experimental and calculated values shown in Ref.~\cite{Rubio_attosecond_response_RT-TDDFT}, which were in the range of $100$ to $120$~meV. Our results also show a substantial time-dependent oscillation, which is absent in the results shown in Ref.~\cite{Rubio_attosecond_response_RT-TDDFT}.
Taking into account the difference of implementation and the difference in the initial value of $U_\eff$ at $t=0$ (our $U_\eff(t=0)$ was 3.697 eV, while that in Ref.~\cite{Rubio_attosecond_response_RT-TDDFT} was $7.82$ eV), we consider the $20$ to $40$ meV deviation of the reduction amplitude to be 
reasonable.

The results obtained by Cazali~\textit{et al.} in Ref.~\cite{Rubio_attosecond_response_RT-TDDFT}, which we reproduced in a different implementation, indicate ACBN0 as a highly promising tool for real-time simulations of correlated systems. This is by virtue of its mean-field-like formulation, where $U_{\eff}$ is calculated only from the static correlation included in the density matrix and the (time-dependent-)KS orbitals.
On the other hand, the fact that the expression of $U_{\eff}$ (Eq.~\eqref{eqn:Ubar_ACBN0}) is not derived from the standard many-body theory makes it difficult to explain and/or predict how the ACBN0 $U_{\eff}$ behaves in dynamical simulations. 
We consider that, for further development of its applications, there has to be a more detailed discussion on the derivation of Eq.~\eqref{eqn:Ubar_ACBN0} and analyses on the behaviors of relevant quantities, such as the renormalized density matrix (Eq.~\eqref{eqn:renormalizedDM_AORep}), in non-equilibrium dynamics.

\subsection{Minimum-tracking linear response, dynamical calculations}\label{subsec:RSL_MTLR_dynamical}
\begin{figure}[ht]
    \centering
    \includegraphics[width=0.8\textwidth]{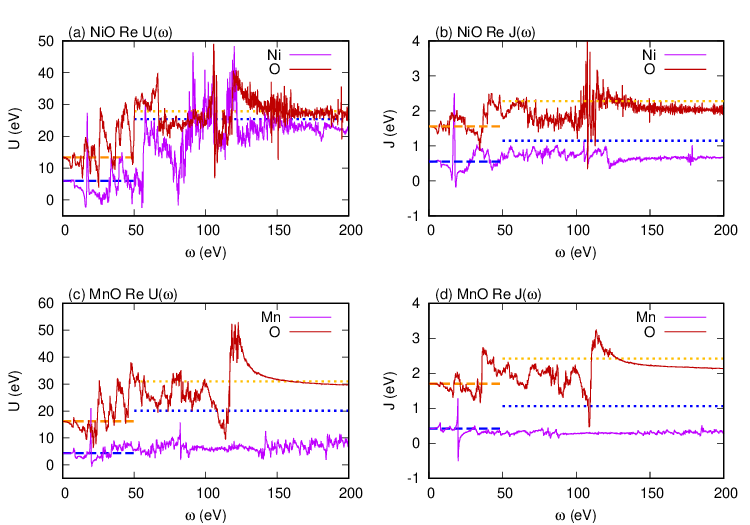}
    \caption{Energy-dependent Hubbard parameters for NiO and MnO. Panels (a) and (b) show the real parts of $U(\omega)$ and $J(\omega)$ for NiO, respectively, whereas (c) and (d) are those for MnO. Purple solid lines show those of metal 3d orbitals, while red solid lines show those for oxygen 2p orbitals. The blue [orange] dashed and dotted horizontal lines show the static linear response calculation results and the unscreened values of the metallic [oxygen] atom, respectively.}
    \label{fig:tdMTLRkp_NiO_MnO}
\end{figure}

We show numerical results of the linear-response-based calculation method for energy-dependent Hubbard parameters we proposed in Sec.~\ref{subsubsec:Extension_of_LR_schemes}.
For benchmark calculations, we calculated the dynamical responses of NiO and MnO, which we worked on in Sec.~\ref{subsec:RSL_MTLR_static}.
We took a $2\times 2\times 2$ supercell of these materials and sampled the Brillouin zone using a $3\times 3\times 3$ Monkhorst-Pack~\cite{Monkhorst-Pack} mesh. We applied the Hubbard parameters derived in the self-consistent calculations in Sec.~\ref{subsec:RSL_MTLR_static}. We set the charge component of the perturbation $\Delta_c$ to $\pm 0.10$ (eV) to get $U(\omega)$ and the spin component $\Delta_s$ to $\pm 0.10$ (eV) to get $J(\omega)$. 
The perturbation was applied to a single atom for each species.
In each simulation, the time-dependent KS orbitals are propagated for $T=24$ fs with step size $\Delta t=8$~as. 
The results of $J(\omega)$ of NiO were obtained from calculations using a step size $\Delta t=5$~as and a propagation time $T=24$ fs. 
The obtained $U(\omega)$ and $J(\omega)$ are shown in Fig.~\ref{fig:tdMTLRkp_NiO_MnO}, purple lines for those of the metallic 3d orbitals, whereas red lines for those of O 2p orbitals.  In addition to (the real parts of) $U(\omega)$ and $J(\omega)$, we also indicated the static calculation results and unscreened values of each parameter (see Appendix~\ref{app:UNscreenedUandJ} for the details) for the metallic [oxygen] atom with blue [orange] dashed and dotted lines, respectively.
We first find that each Hubbard parameter at $\omega = 0$ equals its static value, as expected.
Thus, we can confirm that the new dynamical scheme is a faithful extension of the static minimum-tracking linear-response method in the sense that it reproduces the static results at $\omega=0$.
We next examine the limit of large $\omega$, where the Hubbard parameters are expected to asymptote to unscreened Coulomb interactions. 
In these calculations, we can consider the energy $\hbar \omega \approx 200$ eV to be large enough since relevant single-particle excitations, judged from the imaginary part of $U(\omega)$ or $J(\omega)$ (the imaginary parts of Hubbard parameters are plotted in Appendix~\ref{app:ImaginaryPartOfUandJ}), appear in the energy below $200$~eV.
In Figs.~\ref{fig:tdMTLRkp_NiO_MnO} (a) and (c), we find oxygen $U(\omega)$ (red lines) correctly asymptotes to the unscreened value. 
$U(\omega)$ of Ni 3d (the purple solid line in Fig.~\ref{fig:tdMTLRkp_NiO_MnO}(a)), $J(\omega)$ of O 2p (the red solid lines in Figs.~\ref{fig:tdMTLRkp_NiO_MnO} (b) and (d)) also asymptote to nearby values. On the other hand, we could not get the expected asymptotic behaviors for Mn $U(\omega)$ and $J(\omega)$ of metallic 3d orbitals.
At this point, we do not understand the cause of such unexpected asymptotic behaviors in some of these parameters. 
A possible cause is insufficient numerical accuracy; since both the numerator and denominator in Eqs.~\eqref{eqn:Ubar_tdMTLR} and \eqref{eqn:Jbar_tdMTLR} asymptote $0$ in the limit of large $\omega$, the result may be vulnerable to numerical errors.
We consider that this problem has to be resolved in future studies for the practical application of this energy-dependent Hubbard parameter calculation method.

\section{Summary}\label{sec:Summary}
We implemented \textit{ab initio} calculation methods for Hubbard parameters in the $k$-point sampling RT-TDDFT code in CP2K. 
We worked on a mean-field-like approach, ACBN0, and a linear-response approach, the minimum-tracking linear-response method, where the latter was extended to a dynamical response theory for energy-dependent Hubbard parameters.
In static property calculations, both methods reproduced band gaps and magnetic moments of TMN/Ns in reasonable agreement with experimental results, with a few exceptions. However, we find no clear agreement in the Hubbard parameters obtained with these two methods.
In dynamical applications, the mean-field-like ACBN0 was applied to field-induced dynamics to reproduce the main feature of the calculation results by Cazali~\textit{et al}.
The minimum-tracking linear-response method was extended for the calculation of energy-dependent Hubbard parameters.

%
Our newly proposed calculation scheme of energy-dependent Hubbard parameters has been shown to be consistent with its static counterpart by the values of $U(\omega)$ or $J(\omega)$ at $\omega=0$. On the other hand, we found that some of our numerically obtained results did not exhibit the expected asymptotic behavior as $\omega \to \infty$; while some of the Hubbard parameters asymptoted to the unscreened Coulomb xc interaction, others did not. 
Investigating this issue is left for future studies.

Since our scheme requires multiple RT-TDDFT simulations of supercells with broken symmetry, the efficiency of numerical computation is highly important. In the numerical calculations in this paper, the scalability of the CP2K worked favorably in this respect. To realize calculations with even larger supercells and/or larger number of $\bfk$-point sampling, further optimization of the supercell RT-TDDFT is desirable.    

Our new scheme formally includes the xc effects contained in the xc functional. This is in contrast to cRPA, in which only the Coulombic screening in the random-phase approximation is taken into account. 
The fact that our expression of screening (Eq.~\eqref{eqn:Umm_omega}) does not require strict distinction of the \textit{screening} and \textit{correlated} orbitals may work favorably in materials with strong hybridization among metallic and non-metallic orbitals~\cite{Carta-Timrov_Bridging_cRPA_and_LR_2505.03698v1}. 

We also emphasize the xc-functional dependence of our scheme as an advantage for DFT+U-type applications.  
We recall that in the original static minimum-tracking linear response theory~\cite{Moynihan-ORegan2017self-consistentU,Linscott-ORegan_MTLR_PRB2018}, the (static) Hubbard parameters are designed to remove the self-interaction errors associated with the DFT calculation with the given xc functional. Since the dynamical Hubbard parameters obtained using our scheme equal these static Hubbard parameters at $\omega=0$, and are expected to asymptote to the unscreened Hxc interaction as $\omega\to\infty$, we can expect them to be optimal for DFT+U calculations with the given xc functional both in the low-energy and high-energy limits. 
Applications of such energy-dependent Hubbard parameters in some advanced DFT+U calculation schemes, such as the DFT+U($\omega$)(+V) theory proposed by Vanzini and Marzari~\cite{Marzari_DFT+Uw}, 
might be investigated in the future.

\section*{Data Availability}
The data that supports the findings of this article are not publicly available. The data are available upon reasonable request from the authors.

\section*{Acknowledgment}
We acknowledge funding by the University of Zurich and access to Alps (daint.alps) at the Swiss National Supercomputing Centre, Switzerland (project ID: lp11).

\section*{Author contributions}
K.~H. formulated the new methodology, implemented the computer code, carried out the simulations and analysis, and wrote the initial draft of the paper. S.~L. reviewed the manuscript and secured funding for and supervised the project.
%

\appendix
\section{The linear response method by Cococcioni~\textit{et al.}}
\label{app:CococcioniLRT}
We show calculation results of the linear-response method proposed by Cococcioni~\textit{et al.}~\cite{Cococcioni_DFT+Ulr_PRB2005} implemented in our framework.
We first briefly review the formulation.
In their formulation, one applies a spin-independent perturbation of the form
\begin{align}
    V^{\mathrm{Pert.}}=\lambda^{(\DFTplusUgroup,A)} \calP^{(\DFTplusUgroup,A)}
\end{align} 
and calculates the difference of the occupation number $N^{(\DFTplusUgroup,A')} \equiv \sum_\sigma N_\sigma^{(\DFTplusUgroup,A')}$ at each atomic site $A'$.
Rewriting the combination $(\DFTplusUgroup,A)$ as a single index $I, J, \cdots$, the response function is calculated as
\begin{align}
    \chi^{IJ} = \frac{ \partial N^{I} }{ \partial \lambda^J} 
\end{align}
using $N^{I}$ computed from the converged solution of SCF calculations, whereas the KS response function is calculated as
\begin{align}
    \chi_{KS}^{IJ} = \frac{ \partial N^{I}|_{\mbox{\scriptsize first iteration}} }{ \partial \lambda^J} 
\end{align}
using $N^{I}|_{\mbox{\scriptsize first iteration}}$ computed from the first iteration in the SCF calculation. 
The Hubbard parameter is then obtained as
\begin{align}
    \Ubar^{I}=\left( \chi_{\KS}^{-1} - \chi^{-1} \right)^{II},
    \label{eqn:U_in_dLR}
\end{align}
which is to be interpreted as $U_{\eff}$ in Dudarev's formulation of DFT+U.

We implemented this method into the same framework of our $\bfk$-point sampling RT-TDDFT in CP2K. We used the same projection scheme as we discussed in Sec.~\ref{subsubsec:ProjectionScheme}, which is based on the atomic orbitals of an isolated atom (Eq.~\eqref{eqn:atomicKSO}).
To validate our implementation, we calculated NiO and Cu$_2$O to compare with the results shown in Ref.~\cite{Timrov-Marzari-Cococcioni_DFPTlr}.
The lattice parameter of NiO was set $a=4.1704$~\AA~in the same manner as in Sec.~\ref{subsec:RSL_ACBN0_static}, whereas that for Cu$_2$O was set $a=4.27$~\AA, following Ref.~\cite{Timrov-Marzari-Cococcioni_DFPTlr}. The detailed geometries are shown in the supporting information.

We used the PBE-sol~\cite{PBEsol} xc functional in accordance with Ref.~\cite{Timrov-Marzari-Cococcioni_DFPTlr}. The Hubbard-like on-site interaction was applied only to the metallic $d$-orbitals, and the results were obtained from single-shot/non-self-consistent calculations; i.e., we started DFT+U calculation with a vanishingly small value of $U_{\eff}$ ($U_{\eff}=10^{-7}$ eV) and calculated the linear response of the system only once, as indicated in Ref.~\cite{Timrov-Marzari-Cococcioni_DFPTlr}. Our results are shown in Table~\ref{table:DFTplusUlr} together with corresponding results taken from Ref.~\cite{Timrov-Marzari-Cococcioni_DFPTlr}.
For a technical reason, some of the supercell sizes in our calculations (NiO, $3\times 3\times 2$ and $4\times 4\times 3$) are not the same as the corresponding sizes in Ref.~\cite{Timrov-Marzari-Cococcioni_DFPTlr}, however, we can make a meaningful comparison since the supercell-size dependence of the Hubbard parameters was found to be much smaller than the difference between our results and those in Ref.~\cite{Timrov-Marzari-Cococcioni_DFPTlr}.
From the results of NiO, we find that our results are converging with respect to the supercell size at $3\times3\times2$ or larger; however, the values of U$_{\eff}$ are considerably smaller than those shown in Ref.~\cite{Timrov-Marzari-Cococcioni_DFPTlr}.
On the other hand, for Cu$_2$O, assuming that the supercell size $3\times3\times3$ is large enough also for Cu$_2$O, we find that our obtained value of U$_{\eff}$ is much larger than that in Ref.~\cite{Timrov-Marzari-Cococcioni_DFPTlr}.
We attribute these discrepancies to the difference in 
the pseudopotential, 
the basis set and the projection scheme. 
\begin{table}[ht]
    \centering
    \begin{tabular}{|c|c|c|c|c|}
    \hline
         & \multicolumn{2}{|c|}{ }& \multicolumn{2}{|c|}{U$_{\eff}$ (eV)}\\
         \hline
         & supercell & k-mesh  & this work & \multicolumn{1}{|c|}{Ref.[\cite{Timrov-Marzari-Cococcioni_DFPTlr}]} \\
    \hline
    NiO & 2$\times$2$\times$2 & 6$\times$6$\times$6 & 6.620 & 7.895 \\
    \hline
        & 3$\times$3$\times$2 & 4$\times$4$\times$4 & 6.638 & \\
    \hline
        & 3$\times$3$\times$3 & 4$\times$4$\times$4 &        & 8.146 \\
    \hline
        & 4$\times$4$\times$3 & 3$\times$3$\times$3 &  6.638 & \\
    \hline
        & 4$\times$4$\times$4 & 3$\times$3$\times$3 &        & 8.168 \\
    \hline
    Cu$_2$O & 2$\times$2$\times$2 & 6$\times$6$\times$6 & 16.966 & 11.263 \\
    \hline
     & 3$\times$3$\times$3 & 4$\times$4$\times$4 & 17.698 & 11.287 \\
    \hline     
    \end{tabular}
    \caption{Results Cococcioni's linear response theory calculations. The 4th column shows the value of $U_{\eff}$ (in eV) obtained in our calculations, whereas the 5th column shows those shown in Ref.\cite{Timrov-Marzari-Cococcioni_DFPTlr} }
    \label{table:DFTplusUlr}
\end{table}

\section{Projected density of states (pDOS) calculated using ACBN0} \label{app:ACBN0pDOS}
\begin{figure}[ht]
    \centering
    \includegraphics[width=0.9\linewidth]{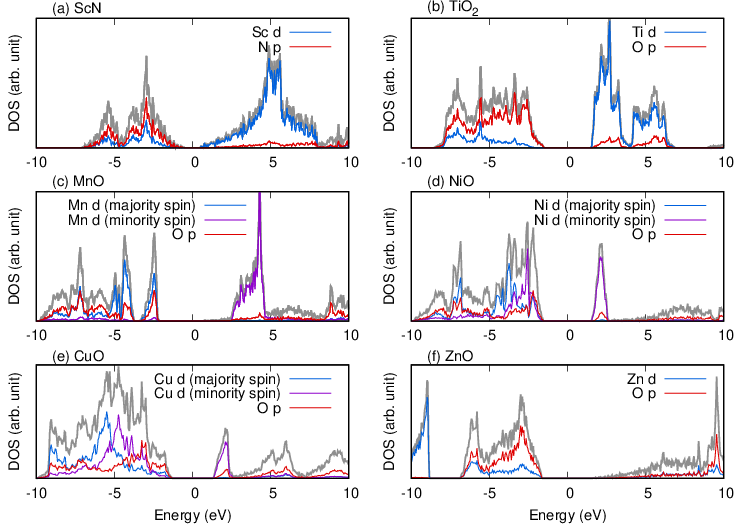}
    \caption{PDOSs of ScN (a), TiO$_2$ (b), MnO (c), NiO (d), CuO (e), and ZnO (f). 
    For non-magnetic systems (a,b,f) the blue lines show the DOS projected on the metallic $3d$ bands, whereas for antiferromagnetic systems (c,d), light-blue and purple lines show the metallic $3d$ bands of majority and minority spins, respectively. The red lines show that projected on oxygen/nitride $2p$ bands. }
    \label{fig:pDOS_ACBN0}
\end{figure}

We show pDOSs of $6$ TMO/N materials we worked on in Sec.~\ref{subsec:RSL_ACBN0_static} to check the reproduction of qualitatively correct band structures of the target materials. 
The pDOSs were calculated based on the L\"owdin population analysis of the Kohn-Sham orbitals. 
In Fig.~\ref{fig:pDOS_ACBN0}, we can confirm relevant features, including $d$-dominant conduction bands (CB) in early TMN and TMO (panels (a) and (b)), a narrow minority-spin $d$-dominant feature of the lowest CB in late TMOs of antiferromagnetic spin structure, MnO, NiO, and CuO (panels (c), (d),  (e)). 
As for the valence bands (VBs) of these materials, which are categorized as charge-transfer insulators, our calculation results show that the highest VBs of these materials are hybrids of oxygen $2p$ and metallic $3d$. As for MnO (panel(c)), this feature differs from the PBE+U calculation by Zunger et al.~\cite{NiOMnOZunger}, which  
shows more 2p-dominant behavior of VB, whereas the GW+U calculations of MnO by Kobayashi \textit{et al.}~\cite{Kobayashi_GW+U} 
show a mixed nature of those VB bands. 
Our pDOS of CuO (Panel (e)) shows Cu 3d bands overlapping with a broad O 2p band, which is in qualitative agreement with the GW calculation results by Wang et al.~\cite{WangCuOGW}. 
Our pDOS of ZnO shows O2p-dominance of the highest VB and small 3d contribution to CB bands, though the dominance of O2p for VB is less pronounced compared to the ACBN0 calculation by Agapito et al.~\cite{Agapito-Nardelli2015ACBN0}.
We consider that the overall qualitative agreement with known properties of each material is satisfactory.

\section{Enforcing symmetry}\label{app:Enforcing_Symmetry}
In large supercell calculations of the (unperturbed) ground states of antiferromagnetic materials, we enforced symmetry with respect to the simultaneous interchange of the A and B lattice sites and the spin projection.
We did this to avoid reaching supercell-size-dependent variational minima with broken symmetry. 
In NiO, for example, taking lattice vectors $\bfa_1$ and $\bfa_2$ perpendicular to the crystal $[111]$ axis, translation of $\bftau = \Ifrac{\bfa_3}{2}$ interchanges sites A and B.
Our target symmetry is therefore written as 
\begin{align}
    \psi_{\lambda\bfk\downarrow}(\bfr) = \psi_{\lambda\bfk\uparrow}(\bfr-\bftau) e^{i\theta_{\lambda\bfk}},
    \label{eqn:AFMsymmetry}
\end{align}
with possible irrelevant phase factor $e^{i\theta_{\lambda\bfk}}$.
We can expand this equation in the AO representation as
\begin{align}
    \sum_\mu \chi_{\mu\bfk}(\bfr)C^{\mu}_{\lambda\bfk\downarrow}
    &= \sum_\nu \chi_{\nu\bfk}(\bfr-\bftau) C^{\nu}_{\lambda\bfk\uparrow} e^{i\theta_{\lambda\bfk}}.
\end{align}
We introduce a new notation of $\bfk$-adapted AO (Eq.~\eqref{eqn:k-adaptedAO}) as
\begin{align}
    \chi_{\mu\bfk}(\bfr) 
    &=\sqrt{\frac{1}{N_L}} \sum_\bfT e^{i\bfk\cdot\bfT} \xi_{[\mu]}(\bfr-\bfR_{[\mu]}-\bfT),
\end{align}
where $[\mu]$ represents the atomic species of the $\mu$th AO and $\bfR_{[\mu]}$ represents the atomic coordinate in the unit cell at the origin.
We can then rewrite Eq.~\eqref{eqn:AFMsymmetry} as
\begin{align}
   \sum_\mu \chi_{\mu\bfk}(\bfr)C^{\mu}_{\lambda\bfk\downarrow}
    &=\sum_\nu \sqrt{\frac{1}{N_L}} \sum_\bfT \xi_{[\nu]}(\bfr-\bfR_{[\nu]} -\bftau)e^{i\bfk\cdot\bfT}C^{\nu}_{\lambda\bfk\uparrow} e^{i\theta_{\lambda\bfk}} \nonumber\\
    &= \sum_\nu \sqrt{\frac{1}{N_L}} \sum_\bfT\xi_{[\nu]}(\bfr-\bfR_{[f(\nu)]}-\mathbf{\Delta}_{[\nu]})e^{i\bfk\cdot\bfT} C^{\nu}_{\lambda\bfk\uparrow} e^{i\theta_{\lambda\bfk}}\nonumber\\
    &= \sum_\nu \chi_{f(\nu)\bfk}(\bfr)e^{-i\bfk\cdot\mathbf{\Delta}_{[\nu]} }C^{\nu}_{\lambda\bfk\uparrow} e^{i\theta_{\lambda\bfk}}
\end{align}
where we introduced index transformation $f(\nu)$ and its associated lattice translation $\mathbf{\Delta}_{[\nu]}$ defined as~\cite{HanasakiLuber}
\begin{align}
    \bfR_{[\nu]} + \bftau = \bfR_{[f(\nu)]} + \mathbf{\Delta}_{[\nu]}
\end{align}
where in the RHS, $\bfR_{[f(\nu)]}$ is taken inside the unit cell at the origin, while a possible lattice translation is absorbed into $\mathbf{\Delta}_{[\nu]}$. 
We can thus obtain the conversion of the orbital coefficient as
\begin{align}
    C^{f(\nu)}_{\lambda\bfk\downarrow} = e^{-i\bfk\cdot\mathbf{\Delta}_\nu} C^{\nu}_{\lambda\bfk\uparrow}.
\end{align}
In numerical calculations, we set the phase factor $e^{i\theta_{\lambda\bfk}}$ to unity.

\section{Calculation of the unscreened Coulomb parameters}
\label{app:UNscreenedUandJ}
We show how we estimated the unscreened Coulomb parameters in Sec.~\ref{subsec:RSL_MTLR_dynamical}.
We performed a separate static minimum-tracking linear response calculation using the Hubbard parameters obtained from self-consistent calculations (Sec.~\ref{subsec:RSL_MTLR_static}).
We perturbed the system with $V^{\mathrm{Pert.}}_\sigma$ (Eq.~\eqref{eqn:Vpert}) on the projected space $(\DFTplusUgroup,A)$ to calculate the induced variance of the orbital occupation numbers as $\delta n^{(\calI,A)}_{m\sigma} \equiv \tr{\rho^{\mathrm{Pert.}}_\sigma |\phi_m^{(\calI,A)}\rangle \langle \phi_m^{(\calI,A)}|} - \tr{\rho^{\mathrm{unpert.}}_\sigma |\phi_m^{(\calI,A)}\rangle \langle \phi_m^{(\calI,A)}|}$, with $\rho_\sigma^{\mathrm{Pert.}}$ and $\rho_\sigma^{\mathrm{unpert.}}$ representing the perturbed and unperturbed density matrices, respectively.
We then constructed a modified density matrix $\widetilde{\rho}_\sigma (\bfr,\bfr')$ having $\delta n^{(\calI,A)}_{m\sigma}$ excess occupation as
\begin{align}
\widetilde{\rho}_\sigma (\bfr,\bfr') =  \rho^{\mathrm{unpert.}}(\bfr,\bfr') + \sum_{m} \phi_m^{(\calI,A)}(\bfr) \delta n_{m\sigma}^{(\calI,A)} \phi_m^{(\calI,A)}(\bfr'),
\end{align}
to calculate the corresponding modified Hartree-xc potential $\widetilde{V}^{Hxc}_\sigma$. We then evaluated its orbital average as
\begin{align}
\overline{ \widetilde{V}}_{\Hxc,\sigma}^{(\calI,A)} = \Tr{\calP^{(\DFTplusUgroup,A)}\widetilde{V}^{\Hxc}_\sigma \calP^{(\DFTplusUgroup,A)}}.
\end{align}
The unscreened $U$ and $J$ were then obtained from the same expressions as Eqs.~\eqref{eqn:Ubar_MTLR} and~\eqref{eqn:Jbar_MTLR} by replacing $\overline{V}^{(\DFTplusUgroup,A)}_{\Hxc,\sigma}$ with $\overline{ \widetilde{V}}_{\Hxc,\sigma}^{(\calI,A)}$.
It is clear that the difference between $\widetilde{V}^{\Hxc}_{\sigma}$ and the unperturbed Hxc potential 
$V^{\Hxc\;\mathrm{unperturb.}}_\sigma$ 
is, up to the first order in $\delta n_{m\sigma}^{(\calI,A)}$,
\begin{align}
\widetilde{V}^{\Hxc}_{\sigma}(\bfr) - V^{\Hxc\;\mathrm{unperturb.}}_{\sigma}(\bfr) \approx \sum_{\tau,m} \int d^3\bfr' f^{\sigma\tau}_{Hxc}(\bfr,\bfr') \left|\phi_m^{(\calI,A)}(\bfr') \right|^2\delta n_{m\tau}^{(\calI,A)},
\label{eqn:diff_VHxc}
\end{align}
and the orbital average of Eq.~\eqref{eqn:diff_VHxc} becomes a weighted average of the unscreened Coulomb interactions.

\section{The imaginary parts of $U(\omega)$ and $J(\omega)$} \label{app:ImaginaryPartOfUandJ}
\begin{figure}[ht]
    \centering
    \includegraphics[width=0.8\linewidth]{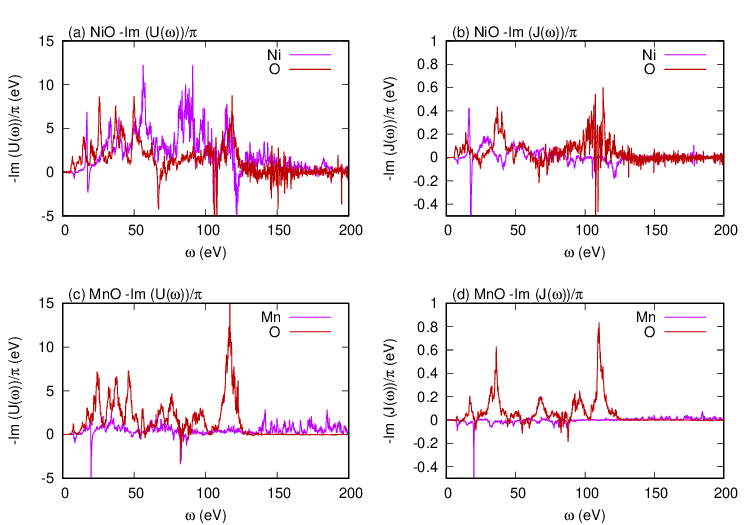}
    \caption{Energy-dependent Hubbard parameters for NiO and MnO. Panels (a) and (b) show the imaginary parts of Hubbard parameters, $-\frac{1}{\pi}\mathrm{Im}U(\omega)$ and $-\frac{1}{\pi}\mathrm{Im}J(\omega)$ for NiO, respectively, whereas (c) and (d) are those for MnO. Purple solid lines show those of metal 3d orbitals, while red solid lines show those for oxygen 2p orbitals.}
    \label{fig:tdMTLRkp_NiO_MnO_Imaginary}
\end{figure}

We show the imaginary parts of the dynamical Hubbard parameters of NiO and MnO calculated using the dynamical extension of the minimum-tracking linear-response method we discussed in Sec.~\ref{subsec:RSL_MTLR_dynamical}.
As indicated in Eq.~\eqref{eqn:Umm_omega}, the imaginary part arises from that of the density response $\chi^{(\mathsf{R})}(\omega)$. The peaks of the imaginary parts of $U(\omega)$ and $J(\omega)$, multiplied by the factor $-\Ifrac{1}{\pi}$, are the spectral peaks of single-particle and collective excitations, weighted by the Hartree-xc kernel $f_{\Hxc}$.
To check such spectral peaks, we plotted, in Fig.~\ref{fig:tdMTLRkp_NiO_MnO_Imaginary}, the imaginary part of the dynamical Hubbard parameters, $-\frac{1}{\pi} \mathrm{Im}U(\omega)$ and $-\frac{1}{\pi}\mathrm{Im}J(\omega)$. 
Since our $U(\omega)$ and $J(\omega)$ are of the retarded form, the spectral function should have, in principle, non-negative values, whereas the $f_{\Hxc}$ kernel may take a negative value.
For simplicity, we assume the negative peaks in Fig.~\ref{fig:tdMTLRkp_NiO_MnO_Imaginary} as arising from possible numerical errors and focus on the positive peaks. From Fig.~\ref{fig:tdMTLRkp_NiO_MnO_Imaginary}, we find that the spectral weights are mostly located in the range $\hbar \omega < 200$~eV. We can therefore expect that the real part of the Hubbard parameters $U(\omega)$ and $J(\omega)$ should reach their unscreened values at around $\hbar\omega\approx 200$~eV.



\bibliography{CP2K_DFTplusU}
\end{document}